\documentclass[a4paper,11pt]{article}
\usepackage{jheppub} 
\usepackage{amsmath}
\usepackage{slashed}
\usepackage{rotating}
\usepackage{array}
\usepackage{orcidlink}
\usepackage{float}

\title{\boldmath Semi-universality of conformal higher-derivative and conformal higher-spin fields}

\affiliation[a]{ Department of Theoretical Physics\\
Tata Institute for Fundamental Research, Mumbai 400005, India.}
\author[a,\orcidlink{0000-0002-1822-2264}]{Jyotirmoy Mukherjee}\emailAdd{jyotirmoy.mukherjee\_119@tifr.res.in}
\author[a,\orcidlink{0000-0002-5553-7003}]{and Pabitra Ray}\emailAdd{raypabitra96@gmail.com}

\abstract{
In this paper, we study thermal partition functions of free exotic conformal field theories, focusing on conformal higher-derivative and conformal higher-spin fields, in the semi-universal limit $|\omega_i|\rightarrow 1$. It was recently conjectured in \cite{Anand:2025mfh} that, in this limit, the thermal partition function develops universal poles in $(1-|\omega_i|)$, while the corresponding residue functions are theory-dependent. We analyze conformal higher-derivative scalar, fermionic, and vector fields in the semi-universal limit. We then extend the study to the Weyl graviton, the Weyl gravitino, and conformal higher-spin fields (CHS) on $S^1_\beta\times S^3$, using both spectral mode-sum and operator-counting methods. In all cases, we find the expected pole structure, with residue functions whose behavior depends on the presence or absence of negative-twist states. For four-dimensional conformal higher-spin fields, we further reproduce the same residue-pole structure from the one-loop partition function of massless higher-spin fields in thermal AdS$_5$. Finally, we show that the semi-universal limit provides a useful diagnostic of negative-twist states, which indicate violations of ANEC-type bounds in these theories, whereas the traditional high-temperature expansion is insensitive to them.
}

\begin{document}
\maketitle
\flushbottom
\section{Introduction}
The partition function of  conformal field theories on $S^1\times S^{d-1}$ is an object of great interest, as it completely characterizes the spectrum of local operators. In general, the partition function of a conformal field theory depends on the theory; however, in certain asymptotic limits, it exhibits universal features.

We consider the thermal partition function on $S^1_\beta\times S^3$, with inverse temperature $\beta$ and two angular chemical potentials $\omega_1$ and $\omega_2$. These chemical potentials are conjugate to the two angular momenta $J_1$ and $J_2$, associated with rotations in the two orthogonal planes when $S^3$ is embedded in $\mathbb{R}^4$. The  canonical partition function is defined by
\begin{equation}\label{partdef}
\begin{split}
Z(\beta,\omega_1,\omega_2)
&=
{\rm{Tr}}\left[e^{
-\beta\left(E-\omega_1 J_1-\omega_2 J_2\right)
}\right]\nonumber\\
&=
{\rm{Tr}}
\left[e^{
-\beta\tau
-\nu_1\,J_1-
\nu_2 J_2
}\right]~.
\end{split}
\end{equation}
We call $\tau=E-J_1-J_2$ the twist and $\nu_i=\beta(1-\omega_i)$ the reduced angular potentials.

It is well known that, from extensivity and dimensional analysis, the logarithm of the partition function of any four-dimensional conformal field theory has the universal high-temperature scaling
$\ln Z\sim \beta^{-3}$,
when $\omega_1=\omega_2=0$. More generally, in $d$ spacetime dimensions, one expects
$\ln Z\sim \beta^{-(d-1)}.$ This result also has a nontrivial generalization to the high temperature, but now with  non-zero angular momenta \cite{Bhattacharyya:2007vs}; see also \cite{Banerjee:2012iz,Jensen:2012jh,Shaghoulian:2015lcn,Benjamin:2023qsc}. In this case, the partition function again takes a universal form up to an undetermined proportionality function of $(\beta,\omega_i)$, which contains the dynamical information of the theory.
As we have seen, the high-temperature regime of the thermal partition function exhibits universal properties. However, partition functions in the regime of chemical potentials, where they take extreme values, remain largely unexplored. Recent studies of grey galaxies \cite{Kim:2023sig,Bajaj:2024utv} have explicitly shown that the angular chemical potentials are also bounded and that the partition functions of CFTs diverge when $|\omega_i|\to 1$ from below.

Motivated by these observations, it has recently been conjectured in \cite{Anand:2025mfh} that the thermal partition function of CFTs on $S_{\beta}^1\times S^{3}$ also exhibits a universal pole structure in the rotational limit $|\omega_i|\to 1$ at a fixed temperature \footnote{The conjecture has also been extended to higher-dimensional CFTs on $S_{\beta}^1\times S^{d-1}$, including the asymmetric limit of the angular momentum. We refer the reader to section 2.5 of \cite{Anand:2025mfh} for a detailed discussion on this.
}. In this limit, the detailed residue functions remain theory-dependent, but the pole structure itself is expected to be universal. In particular, in the limit 
\begin{equation}\label{limit}
    \nu_i\rightarrow 0~, \qquad \text{where} \qquad \nu_i=\beta(1-\omega_i)~,\qquad \frac{\nu_1}{\nu_2}\qquad\text{fixed}~,
\end{equation}
the partition function of conformal field theory is given by \cite{Anand:2025mfh} \footnote{See also the recent developments in this direction \cite{Komargodski:2026ain}.}
\begin{equation}
    \ln Z=\frac{\beta^2\, h(\beta)}{\nu_1\nu_2}+O\left(\frac{1}{\nu_i}\right)+\cdots~.
\end{equation}
The conjecture is formulated based on the general framework of equilibrium partition functions \cite{Banerjee:2012iz,Jensen:2012jh}, sometimes referred to as `thermal effective field theory' \cite{Benjamin:2023qsc}.
In this paper, we study the partition function of conformal higher-derivative and conformal higher-spin fields on $S^1_{\beta}\times S^3$ in the limit \eqref{limit}, using both spectral mode-sum and the operator counting approach. We test the conjecture of \cite{Anand:2025mfh} for these theories and study the properties of the residue functions in the semi-universal limit.
\begin{table}[H]
\centering
\renewcommand{\arraystretch}{1.8}
\setlength{\tabcolsep}{10pt}
\begin{tabular}{|>{\centering\arraybackslash}m{0.14\textwidth}|
                >{\centering\arraybackslash}m{0.74\textwidth}|}
\hline
Field  & Leading residue $h(\beta)$ \\
\hline

$\phi$
&
$\displaystyle 
\frac{1}{\beta^2}
\sum_{k=1}^{\infty}
\frac{1}{k^3}\frac{1}{2\sinh(k\beta)}
$
\\[4pt]
\hline

$\phi_{4\partial}$
&
$\displaystyle 
\frac{1}{\beta^2}
\sum_{k=1}^{\infty}
\frac{\coth(k\beta)}{k^3}
$
\\[4pt]
\hline

$\psi$
&
$\displaystyle 
\frac{1}{\beta^2}
\sum_{k=1}^{\infty}
\frac{(-1)^{k+1}}{k^3}
\frac{1}{\sinh\left(\frac{k\beta}{2}\right)}
$
\\[4pt]
\hline

$\psi_{3\partial}$
&
$\displaystyle 
\frac{1}{\beta^2}
\sum_{k=1}^{\infty}
\frac{(-1)^{k+1}}{k^3}
\frac{\left(1+2\cosh(k\beta)\right)}{\sinh\left(\frac{k\beta}{2}\right)}
$
\\[4pt]
\hline

$A_\mu$
&
$\displaystyle 
\frac{1}{\beta^2}
\sum_{k=1}^{\infty}
\frac{\coth(k\beta)}{k^3}
$
\\[4pt]
\hline
$A^{6d}_{M,\,4\partial}$ & $\displaystyle  \frac{1}{\beta^{2}}\sum_{k=1}^{\infty}\frac{2\coth^{2}(k\beta)}{k^{4}}$
\\[4pt]
\hline

$\psi^{\rm Weyl}_\mu$
&
$\frac{1}{\beta^2}\sum_{k=1}^{\infty}\frac{(-1)^{k+1}}{k^3}\frac{4\cosh^2\left(k\beta\right)}{\sinh\left(\frac{k\beta}{2}\right)}$
\\[4pt]
\hline

$h^{\rm Weyl}_{\mu\nu}$
&
$\frac{1}{\beta^2}\sum_{k=1}^{\infty}\frac{\coth\left(k\beta\right)\left(3+4\sinh^2\left(k\beta\right)\right)}{k^3}$
\\[4pt]
\hline

$\Phi_{\mu_1\mu_2\cdots\mu_s}$
&
$\frac{1}{\beta^2}\sum_{k=1}^{\infty}\frac{\sinh\left(s\,k\beta\right)\sinh\left((s+1)\,k\beta\right)}{2k^3\sinh^3\left(k\beta\right)}$
\\[4pt]
\hline

\end{tabular}
\caption{
Residue functions in the semi-universal limit. For the \(4d\) theories,
\(\ln Z=\beta^2 h(\beta)/(\nu_1\nu_2)+\cdots\), with
\(\nu_i=\beta(1-\omega_i)\). For the \(6d\) vector
\(A^{6d}_{M,\,4\partial}\), we define
\(\ln Z=\beta^2 h^{6d}_1(\beta)/(\nu_1\nu_2\nu_3)+\cdots\).
}
\label{tab:hbeta-fields}
\end{table}
In the table above, we list the residue functions of the leading singularities of the partition functions for the conformal fields studied in this paper. Throughout the paper, we focus on the $\omega_i\to 1$ limit; an analogous analysis can be carried out for the $\omega_i\to -1$ limit as well.

This paper further explores the connection between the semi-universal limit and negative-twist states, which are associated with violations of ANEC-type bounds. It was observed in \cite{Anand:2025mfh}, for some exotic free theories \cite{Loganayagam:2012zg}, that the semi-universal limit detects the presence of negative-twist states, whereas the
traditional high-temperature expansion of the partition function is insensitive to them. In this paper, we explore this direction further by studying conformal higher-spin fields
on $S^1_{\beta}\times S^3$ in the semi-universal limit \eqref{limit}.
It turns out that, as summarized in table \ref{tab2}, the partition function of a conformal higher spin field of spin-$s$ on $S_{\beta}^1\times S^3$ exhibits two different behaviors depending on the order in which the limits are taken: first, the small-$\beta$ expansion, and second, the small-$\nu_i$ expansion.

In the small-$\beta$ expansion, the partition function is completely well-behaved
\begin{equation}
\ln Z^{\rm CHS}_s\approx
\frac{1}{\nu_1\nu_2}
\left[
\frac{s(s+1)}{2\beta}
\sum_{k=1}^{\infty}\frac{1}{k^4}
+O(\beta)
\right]
+\cdots .
\end{equation}
However, in the small-$\nu_i$ expansion (with fixed $\beta$) gives \begin{equation}\label{smallnul}
\ln Z^{\rm CHS}_s= \frac{1}{\nu_1\nu_2}
\sum_{k=1}^{\infty}
\frac{
\sinh(sk\beta)\sinh((s+1)k\beta)
}
{2k^3\sinh^3(k\beta)}+O\left(\frac{1}{\nu_i}\right)+\cdots~.
\end{equation}
The large $k$-terms in the sum of \eqref{smallnul} grow like $\frac{e^{2(s-1)k\beta}}{k^3}$.
Therefore, in the semi-universal limit, the fixed-$\beta$ series is convergent only when $s\leq 1$. Therefore, all conformal higher-spin fields with $s\geq 2$, which are expected to violate the ANEC bound, also exhibit pathological behavior in the semi-universal limit, although they are perfectly well-behaved in the traditional high temperature expansion. Thus, the semi-universal limit provides a useful diagnostic of the violation of ANEC-type bounds \cite{Cordova:2017dhq}, whereas the traditional high temperature expansion is insensitive to it.

Based on this analysis, we organize the theories studied in this paper into the following four classes:
\begin{itemize}
    \item \textit{Class 1:}  These theories obey the usual unitarity bounds and do not contain any negative-twist states. Examples include the two-derivative conformal scalar, the massless Dirac fermion, and the Maxwell field. Their thermal partition functions are well behaved, and they obey the expected semi-universal structure.

    \item \textit{Class 2:} Theories that violate the naive unitarity bounds but do not contain negative-twist states. These theories are non-unitary in the traditional sense, but nevertheless have well-behaved partition functions in the semi-universal limit. The four-derivative conformal scalar and the conformal higher derivative vector provide important examples of this class.

    \item \textit{Class 3:} Theories that contain negative-twist states even though they do not obviously violate the naive unitarity bounds. Such theories would be pathological in the semi-universal limit but will obey the naive unitarity bounds. This class is kept as a logically distinct possibility but is not populated by our main examples.

    \item \textit{Class 4:} Theories that both violate the naive unitarity bounds and contain negative-twist states. These theories show pathological behavior in the semi-universal limit. Examples include the three-derivative conformal fermion, Weyl gravitino, the Weyl graviton, and, more generally, conformal higher-spin fields with \(s>1\).
\end{itemize}
We find that theories in \textit{Class 1} and \textit{Class 2} have well-behaved partition functions and obey the expected semi-universal structure, whereas theories in \textit{Class 3} and \textit{Class 4} exhibit pathological behavior in the semi-universal limit.

Let us now outline the organization of the paper. Throughout the paper, we derive the relevant partition functions using both the mode-sum approach and the operator-counting method and study their properties in the semi-universal limit. In section \ref{sc}, as a warm-up, we begin with the two-derivative scalar field on $S^1_\beta\times S^3$, and then generalize the analysis to higher-derivative scalar fields in the semi-universal limit. In section \ref{fr1}, we study the partition functions of the massless Dirac fermion and conformal higher-derivative fermions in the same limit. In section \ref{vc}, we extend the analysis to the Maxwell field and conformal higher-derivative vector fields, obtaining their partition functions in the semi-universal limit. In sections  \ref{grt} and \ref{grav}, we apply the same framework to the  Weyl gravitino and Weyl graviton, respectively. In section \ref{chsbdy}, we derive the conformal higher-spin partition function on the boundary, and in section \ref{bulkchs} we reproduce the same semi-universal structure from the bulk one-loop partition function. Finally, in section \ref{ansc}, we explore the possible connection between semi-universality and the ANEC \cite{Hofman:2008ar}.

\section{Semi-universality of a conformally coupled scalar on $S^1_\beta \times S^3$}\label{sc}
Our main interest is to study partition functions in the semi-universal
limit \eqref{limit}. In the following sections, using both the mode-sum
approach and the operator-counting method, we will compute the partition
functions of conformal higher-derivative fields and conformal higher-spin
fields in this limit.

  As a warm-up, we first rederive the partition function of a conformally
coupled scalar on \(S^1_{\beta}\times S^3\) using the mode-sum approach.
The same result can also be obtained by the operator-counting method via
the state-operator map; see, for example, section 5 of
\cite{Anand:2025mfh}.
\subsection{Two-derivative conformal scalar}
The theory of a conformally coupled two-derivative scalar belongs to the \textit{Class 1}, as mentioned in the introduction. It obeys the usual
unitarity bounds and contains no negative-twist states.

The Euclidean action of  a conformally coupled scalar field on
$
S^1_{\beta}\times S^3,
$ is given by
\begin{equation}
S_0
=
\frac12 \int d^4x \sqrt{g}\,
\left[
(\partial\phi)^2+\frac{R}{6}\phi^2
\right]~.
\end{equation}
The logarithm of the partition function can be written as
\begin{equation}\label{propertime}
\ln Z_0
=
-\frac12 \ln \det {\mathcal O}_0~,
\qquad
{\mathcal O}_0\equiv-\partial_0^2-\nabla_{S^3}^2+\frac{R}{6}~.
\end{equation}
Here $R$ is the Ricci scalar on $S^3$.

It is convenient to rewrite the determinant in Schwinger proper-time form,
\begin{equation}\label{scr}
\ln Z_0
=
\frac{1}{2} \int_0^\infty \frac{ds}{s}\,
{\rm Tr}\left(e^{-s{\mathcal O}_0}\right)~.
\end{equation}
To incorporate the angular chemical potentials $\omega_1$ and $\omega_2$, associated with the two commuting rotations of $S^3$, we impose twisted periodic boundary conditions along the thermal circle in the following way:
\begin{equation}\label{twistedbcs}
\phi(\tau+\beta,\Omega)
=
e^{\,\beta\,\omega_1 J_1+\beta\,\omega_2 J_2}\,
\phi(\tau,\Omega)~.
\end{equation}
Here $\Omega$ denotes the angular coordinate on $S^3$, and $J_1$ and $J_2$ are the generators of rotations in two orthogonal planes of $\mathbb{R}^4$, in which $S^3$ is embedded, providing a convenient choice of Cartan generators of the isometry group $SO(4)$. Using the decomposition
\begin{equation}
SO(4)\simeq SU(2)_L\times SU(2)_R~,
\end{equation}
it is convenient to define the left and right generator
\begin{equation}
J_L^3=\frac{J_1+J_2}{2}~,
\qquad
J_R^3=\frac{J_1-J_2}{2}~.
\end{equation}
Scalar harmonics $Y_{\ell;m_L,m_R}$ on $S^3$ belong to the representation
\begin{equation}
\left(\frac{\ell}{2},\frac{\ell}{2}\right)~,
\qquad
\ell=0,1,2,\cdots
\end{equation}
with quantum numbers
\begin{equation}
m_L,m_R=-\frac{\ell}{2},-\frac{\ell}{2}+1,\cdots,\frac{\ell}{2}
\end{equation}
where $m_L$ and $m_R$ are the eigenvalues of $J_L^3$ and $J_R^3$ respectively, i.e,
\begin{equation}
J_L^3 Y_{\ell;m_L,m_R}=m_L Y_{\ell;m_L,m_R}~,
\qquad
J_R^3 Y_{\ell;m_L,m_R}=m_R Y_{\ell;m_L,m_R}~.
\end{equation}
To compute the partition function, we sum over the eigenspectrum of the kinetic operator. We therefore begin by writing the mode expansion of the scalar field on $S^1_{\beta}\times S^3$:
\begin{equation}
\phi(\tau,\Omega)
=
\sum_{n\in\mathbb Z}
\sum_{\ell=0}^{\infty}\sum_{m_L,m_R}
c_{n;\ell;m_L,m_R}\,
e^{i\omega_n\tau}
Y_{\ell;m_L,m_R}(\Omega)~.
\end{equation}
The twisted boundary condition in \eqref{twistedbcs} implies
that the shifted Matsubara frequencies are
\begin{equation}\label{Matsubara}
\omega_n
=
\frac{2\pi n}{\beta}
-
i\mu~,
\qquad n\in\mathbb Z
\end{equation}
where we have introduced the parameter $\mu$ as 
\begin{equation}\label{mudef}
\mu(m_L,m_R)=(\omega_1+\omega_2)m_L+(\omega_1-\omega_2)m_R~.
\end{equation}
The complete set of eigenvalues\footnote{For a conformally coupled scalar on the unit $S^3$,
\begin{equation}
-\nabla_{S^3}^2+\frac{R}{6}
\quad\longrightarrow\quad
(\ell+1)^2.
\end{equation}} of ${\mathcal O}_0$ are given by :
\begin{equation}
\lambda^{(0)}_{n,\ell}
=\omega^2_n
+
(\ell+1)^2~.
\end{equation}
With the full set of eigenmodes, let us now compute the partition function of a conformally coupled scalar field. Substituting the eigenspectrum into the proper-time representation given in \eqref{scr}, we find
\begin{align}
\ln Z_0
=
\frac12\int_0^\infty \frac{ds}{s}
\sum_{n\in\mathbb{Z}}\sum_{\ell=0}^{\infty}\sum_{m_L,m_R}
e^{
-s\omega_n^2-s(\ell+1)^2}~.
\end{align}
The sum over Matsubara modes $n$ can be performed naturally using the Poisson
resummation formula
\begin{equation}\label{poisson}
\sum_{n\in\mathbb{Z}}e^{-s\omega_n^2}=\sum_{n\in\mathbb Z}
e^{-s\left(\frac{2\pi n}{\beta}-i\mu\right)^2}
=
\frac{\beta}{\sqrt{4\pi s}}
\sum_{k\in\mathbb Z}
e^{-\frac{\beta^2k^2}{4s}}\,e^{k\beta\mu}~.
\end{equation}
Therefore, free energy can be written as
\begin{align}
\ln Z_0
&=
\frac{\beta}{2\sqrt{4\pi}}
\sum_{k\in\mathbb Z}\sum_{\ell=0}^{\infty}
\sum_{m_L,m_R}e^{k\beta\mu}
\int_0^\infty ds\, s^{-3/2}
e^{-s(\ell+1)^2-\frac{\beta^2k^2}{4s}}~.
\end{align}
The $k=0$ term is the zero-winding contribution. It contributes to the Casimir piece which is proportional to $\beta$, which we drop from the thermal partition function. For $k\neq 0$, we perform the integral over $s$, and obtain
\begin{equation}
\int_0^\infty ds\, s^{-3/2}
e^{-s(\ell+1)^2-\frac{\beta^2k^2}{4s}}
=\frac{2\sqrt{\pi}}{\beta |k|}
e^{-\beta |k|(\ell+1)}~.
\end{equation}

Thus the temperature-dependent part of the logarithm of the partition function is given by
\begin{equation}\label{logzsccas}
\begin{split}
\ln Z_0
&=
\frac{\beta}{2\sqrt{4\pi}}
\sum_{k\in\mathbb Z\setminus\{0\}}\sum_{\ell=0}^{\infty}
\frac{2\sqrt{\pi}}{\beta|k|}
e^{-\beta|k|(\ell+1)}\sum_{m_L,m_R}e^{k\beta\mu}\\
&=
\frac{1}{2}
\sum_{k\in\mathbb Z\setminus\{0\}}
\sum_{\ell=0}^{\infty}\frac{1}{|k|}
e^{-\beta|k|(\ell+1)}\sum_{m_L,m_R}e^{k\beta\mu}~.
\end{split}
\end{equation}
Now, combining the $k>0$ and $k<0$ branches, we obtain
\begin{align}
\ln Z_0
&=
\sum_{k=1}^{\infty}\sum_{\ell=0}^{\infty}
\frac{1}{k}
e^{-k\beta(\ell+1)}
\sum_{m_L,m_R}\cosh(k\beta\mu)~,
\end{align}
where we have used the following property of $\mu(m_L,m_R)$
\begin{equation}
\mu(-m_L,-m_R)=-\mu(m_L,m_R)~.    
\end{equation}
The sum over $m_L$ and $m_R$ can be organized nicely in the following way:
 \begin{align}
 \sum_{m_L,m_R}\cosh(k\beta\mu)
 &=
 \frac12
 \sum_{m_L,m_R}
 \left[
 e^{k\beta\mu(m_L,m_R)}
 +
 e^{-k\beta\mu(m_L,m_R)}
 \right]
 \nonumber\\
 &=
 \frac12
 \left[
 \sum_{m_L,m_R}
 e^{k\beta\mu(m_L,m_R)}
 +
 \sum_{m_L,m_R}
 e^{k\beta\mu(-m_L,-m_R)}
 \right]
 \nonumber\\
 &=
 \sum_{m_L,m_R}
 e^{k\beta\mu(m_L,m_R)} .
 \end{align}
 Therefore, the partition function can be written in the following way:
 \begin{align}
 \ln Z_0
 &=
 \sum_{\ell=0}^{\infty}
 \sum_{k=1}^{\infty}
 \frac{1}{k}
 e^{-k\beta(\ell+1)}
 \sum_{m_L,m_R}
 e^{k\beta\mu(m_L,m_R)} \nonumber\\
 &=\sum_{\ell=0}^{\infty}
 \sum_{k=1}^{\infty}
 \frac{1}{k}
 e^{-k\beta(\ell+1)}\, \chi_\ell(k;\omega_1,\omega_2)
 \end{align}
 where we identify, $\chi_{\ell}(k;\omega_1,\omega_2)$ as a product of two $SU(2)$ characters with the potentials $\left(\omega_1+\omega_2\right)$ and $\left(\omega_1-\omega_2\right)$.
 \begin{equation}\label{chars}
 \chi_\ell(k;\omega_1,\omega_2)
 =
 \frac{
 \sinh\!\left(\frac{k\beta(\ell+1)(\omega_1+\omega_2)}{2}\right)
 }{
 \sinh\!\left(\frac{k\beta(\omega_1+\omega_2)}{2}\right)
 }
 \frac{
 \sinh\!\left(\frac{k\beta(\ell+1)(\omega_1-\omega_2)}{2}\right)
 }{
 \sinh\!\left(\frac{k\beta(\omega_1-\omega_2)}{2}\right)
 }.
 \end{equation}
We can now perform the geometric sum over $\ell$, yielding the full partition function, which can be expressed as a bosonic plethystic sum
\begin{align}
\ln Z_0=\sum_{k=1}^{\infty}\frac1k\,z_{0}(k)~.
\end{align}
Here $z_0(k)$, which we identify the single-particle partition function is given by
\begin{equation}\label{zspsc}
z_{0}(k)
=
\frac{e^{-\beta k}\big(1-e^{-2k\beta}\big)}
{(1-e^{-k\beta(1-\omega_1)})(1-e^{-k\beta(1+\omega_1)})
 (1-e^{-k\beta(1-\omega_2)})(1-e^{-k\beta(1+\omega_2)})}~.
\end{equation}

This single-particle partition function can be understood very easily using the operator counting method. The scalar primary has conformal weight $\Delta_{\phi}=1$, and descendant states are generated by acting derivatives on the primary. Each derivative raises the energy by one unit and therefore contributes to a factor of $q$ where $q\equiv e^{-\beta k}$. The four derivatives transform as the vector representation of $SO(4)$ and carry weights $e^{\pm \omega_1}$ and $e^{\pm \omega_2}$. Thus, the unrestricted descendant module contributes
$$\frac{1}{\big(1-q x_1\big)\big(1-qx^{-1}_1\big)\big(1-q x_2\big)\big(1-q x^{-1}_2\big)}~.$$
We will also have to remove the null states corresponding to the equation of motion $D^2\phi=0$
\begin{align}
    z_{0}&=\frac{q\big(1-q^2\big)}
{\big(1-q x_1\big)\big(1-qx^{-1}_1\big)\big(1-q x_2\big)\big(1-q x^{-1}_2\big)}~,
\end{align}
which agrees with \eqref{zspsc}.\\

We will now study the partition function in the semi-universal limit \eqref{limit}.
In this limit, the leading divergent part of the partition function is
\begin{equation}
\ln Z_0
=
\frac{\beta^2 h_{0}(\beta)}{\nu_1\nu_2}
+
O\!\left(\frac{1}{\nu_i}\right)
+\cdots ,
\end{equation}
where
\begin{equation}\label{hbetasc2}
h_{0}(\beta)
=
\frac{1}{\beta^2}
\sum_{k=1}^{\infty}
\frac{e^{-k\beta}}
{k^3(1-e^{-2k\beta})}
=
\frac{1}{2\beta^2}
\sum_{k=1}^{\infty}
\frac{1}{k^3\sinh(k\beta)}~.
\end{equation}
It was shown in \cite{Anand:2025mfh} that the partition function of a free massless scalar on
$S^1_{\beta}\times S^3$ exhibits a semi-universal divergent structure in the
limit \eqref{limit}, with the residue function given by \eqref{hbetasc2}. In this
work, we test this conjectured structure primarily for conformal
higher-derivative and conformal higher-spin fields. 
\subsection{Four-derivative conformal scalar}
The theory of four-derivative conformally coupled scalar was studied in the context of the one-loop beta function of conformal supergravity \cite{Fradkin:1981jc}.
It  belongs to \textit{Class 2} as mentioned in the introduction. It violates the naive unitarity bound but does not contain any negative twist.

The Euclidean action for the four-derivative conformally coupled scalar is given by \cite{Fradkin:1981jc}
\begin{equation}
S_\Phi
=
\frac12 \int d^4x \sqrt{g}\,
\Phi\,\mathcal O_\Phi\,\Phi~,
\end{equation}
where the fourth-order conformally covariant scalar operator is
\begin{equation}
\mathcal O_\Phi
\equiv
D^4
+
2\left(R_{\mu\nu}-\frac13 R g_{\mu\nu}\right)
D^\mu D^\nu~.
\end{equation}
In the background $S^1_\beta\times S^3$, this operator simplifies to
\begin{equation}
\mathcal O_\Phi
=
D^4-4\partial_0^2~,
\qquad
D^2\equiv\partial_0^2+\nabla^2_{S^3}~.
\end{equation}
Here $D^2$ denotes the Laplacian on $S^1_\beta\times S^3$, while
$\nabla^2_{S^3}$ denotes the Laplacian on $S^3$.
For the fourth-order conformal scalar, the kinetic operator on
$S^1_\beta\times S^3$ factorizes as
\begin{equation}
\mathcal O_\Phi=\big[(\partial_0-1)^2+\nabla^2_{S^3}-1\big]\big[(\partial_0+1)^2+\nabla^2_{S^3}-1\big]~.
\end{equation}
Turning on the same angular chemical potentials for the two planes of rotation,
as in the two-derivative scalar case, the Matsubara frequencies are shifted to \eqref{Matsubara} and $\mu$ is defined in \eqref{mudef}. Therefore the full eigenvalues of $\mathcal O_\Phi$ are
\begin{equation}
\lambda^{(\Phi)}_{n,\ell}
=
\bigl(\omega_n^2+\ell^2\bigr)
\bigl(\omega_n^2+(\ell+2)^2\bigr)~.
\end{equation}

Using the Schwinger proper-time representation of the partition function
\begin{equation}
\begin{split}
\ln Z_{\Phi}
&=
-\frac12\ln\det {\mathcal O}_\Phi\nonumber\\
&
=
\frac12\int_0^\infty \frac{ds}{s}
\sum_{n\in\mathbb{Z}}\sum_{\ell=0}^{\infty}\sum_{m_L,m_R}
\left[
e^{-s(\omega_n^2+\ell^2)}
+
e^{-s(\omega_n^2+(\ell+2)^2)}
\right]~.
\end{split}
\end{equation}
Repeating the same Poisson resummation technique, as in the two-derivative case, and
dropping the zero-winding Casimir contribution, we obtain the thermal partition function
\begin{align}
\ln Z_{\Phi}
&=
\sum_{m_L,m_R}\sum_{k=1}^{\infty}\sum_{\ell=0}^{\infty}
\frac1k
\left(
e^{-k\beta\ell}
+
e^{-k\beta(\ell+2)}
\right) \cosh(\mu k\beta)~,
\end{align}
where $\mu$ is defined in \eqref{mudef} and it has an explicit dependence on $m_L$ and $m_R$.
As before, using the symmetry $(m_L, m_R)\to (-m_L, -m_R)$, we find
\begin{equation}
\mu(-m_L,-m_R)=-\mu(m_L,m_R)    
\end{equation}
we can write the sum over $(m_L,m_R)$ on $\cosh(k\beta\mu)$ by the corresponding character sum,
\begin{equation}
\sum_{m_L,m_R}\cosh(k\beta\mu)
= \chi_\ell(k;\omega_1,\omega_2)~,
\end{equation}
where $\chi_\ell(k;\omega_1,\omega_2)$ is the same as
defined in the two-derivative case \label{chars}. Hence
\begin{align}
\ln Z_\Phi
&=
\sum_{k=1}^{\infty}\sum_{\ell=0}^{\infty}\frac1k
\left(
e^{-k\beta\ell}
+
e^{-k\beta(\ell+2)}
\right)
 \chi_\ell(k;\omega_1,\omega_2)~.
\end{align}
Writing $q=e^{-k\beta}$, the remaining $\ell$-sum can be performed exactly,
\begin{equation}
\sum_{\ell=0}^{\infty}
\left(q^\ell+q^{\ell+2}\right)
 \chi_\ell(k;\omega_1,\omega_2)
=
\frac{1-q^4}
{(1-e^{-k\beta(1-\omega_1)})(1-e^{-k\beta(1+\omega_1)})
 (1-e^{-k\beta(1-\omega_2)})(1-e^{-k\beta(1+\omega_2)})}~.
\end{equation}
Therefore, the full multiparticle partition function is
\begin{align}
\ln Z_\Phi
&=
\sum_{k=1}^{\infty}\frac1k\,z_\Phi(k).
\end{align}
Thus we identify the refined single-particle partition function as
\begin{equation}\label{sp4der}
z_\Phi(k)
=
\frac{1-e^{-4k\beta}}
{(1-e^{-k\beta(1-\omega_1)})(1-e^{-k\beta(1+\omega_1)})
 (1-e^{-k\beta(1-\omega_2)})(1-e^{-k\beta(1+\omega_2)})}.
\end{equation}

The partition function of the four-derivative conformal scalar can also be
understood directly using the operator counting method, from the state-operator map. The scalar primary $\Phi$ has
dimension $\Delta_\Phi=0$. Descendant states are generated by acting 
on the derivatives on the primary $\Phi$. Each derivative raises the energy by one unit, and therefore
contributes a factor of $q=e^{-k\beta}$. The four derivatives transform as the
vector representation of $SO(4)$ and carry angular weights
$e^{\pm \omega_1}$ and $e^{\pm \omega_2}$. Thus, the unrestricted descendant
module gives
\begin{equation}
\frac{1}
{(1-qx_1)(1-qx^{-1}_1)
 (1-qx_2)(1-qx^{-1}_2)},
\end{equation}
where $x_i=e^{k\beta\omega_i}$.

For the four-derivative scalar, the equation of motion is
\begin{equation}
D^4\Phi=0.
\end{equation}
Therefore, the descendants generated by the equation of motion $D^4\Phi$ must be removed. Since
$D^4\Phi$ lies four derivative levels above the primary, this subtracts
the same descendant module weighted by $q^4$. Therefore, the refined single-particle partition function is
\begin{equation}
z_\Phi(q;\omega_1,\omega_2)
=
\frac{1-q^4}
{(1-qx_1)(1-qx^{-1}_1)
 (1-qx_2)(1-qx^{-1}_2)},
\end{equation}
which agrees with \eqref{sp4der}.


Let us now study the four-derivative conformally coupled scalar field in the limit \eqref{limit}, 
\begin{align}
\ln Z_\Phi
&=
\frac{1}{\nu_1\nu_2}
\sum_{k=1}^{\infty}
\frac{1}{k^3}
\frac{1+e^{-2k\beta}}
{1-e^{-2k\beta}}
+
O\!\left(\frac{1}{\nu_i}\right)
\nonumber\\
&=
\frac{1}{\nu_1\nu_2}
\sum_{k=1}^{\infty}
\frac{\coth(k\beta)}{k^3}
+
O\!\left(\frac{1}{\nu_i}\right).
\end{align}
Therefore, writing down the partition function in the following way
\begin{equation}
\ln Z_\Phi
=
\frac{\beta^2 h_{\Phi}(\beta)}{\nu_1\nu_2}
+
O\!\left(\frac{1}{\nu_i}\right)
+\cdots ,
\end{equation}
we can read off the residue function of the leading singularity in the semi-universal limit \eqref{limit}
\begin{equation}\label{h4derphi}
h_{\Phi}(\beta)
=
\frac{1}{\beta^2}
\sum_{k=1}^{\infty}
\frac{\coth(k\beta)}{k^3}.
\end{equation}
Note that the residue function $h_{\Phi}(\beta)$ is still well defined, although the theory has more than two-derivatives in the kinetic term. The summand is not growing with large $k$. It is known that this theory violates unitarity but has well defined partition functions; in particular, it obeys semi-universality. We will discuss more about the relationship with the ANEC-type bound in section \ref{ansc}. 
\section{Semi-universality of  Weyl invariant fermions on $S^1_{\beta}\times S^3$}\label{fr1}
In this section, we derive the partition function of a massless Dirac fermion and a higher-derivative Weyl fermion on $S^1_{\beta}\times S^3$ in the semi-universal limit. 
\subsection{Massless Dirac fermion}\label{dirac}
The massless Dirac field belongs to \textit{Class 1}: it obeys unitarity and does not contain negative twist states. 
The action of the massless Dirac fermion is given by
\begin{align}
S_{\frac{1}{2}}
&=
i\int d^4x\sqrt{g}\,\bar\psi\, \gamma^\mu D_\mu\psi,
\\\gamma^\mu=e^\mu_a\gamma^a,\qquad
D_\mu
&=
\partial_\mu+\frac{1}{2}(\omega_{\mu})^{ab}\sigma_{ab},
\qquad
\sigma_{ab}=\frac{1}{4}[\gamma_a,\gamma_b] .
\end{align}
Here $e^{\mu}_a$ are the vielbeins, and $(\omega_{\mu})^{ab}$ is the usual spin connection. 
The partition function of the massless Dirac fermion, can be written as
\begin{align}
Z_{\frac{1}{2}}\label{partD}
&=
\det\big( i\slashed D\big)
=
\left[\det \left(-\slashed D^{\,2}\right)\right]^{\frac{1}{2}},
\end{align}
where the operator $  \slashed D^{\,2}$ is given by
\begin{align}
    \slashed D^{\,2}
&=\partial^2_0+\slashed\nabla^2_{S^3}~,\qquad
\mathbf{\slashed \nabla}_{S^3}=\gamma^i\nabla_i~.
\end{align}
We will now compute the eigenspectrum of the kinetic operator of the Dirac fermion on $S^1_{\beta}\times S^3$. In order to do that, we need to understand the mode functions of Dirac fermion on $S^3$. 
It is well known that, the spinor harmonics $Y^+_{\ell;m^+_L,m^+_R}$ on $S^3$ belong to the representation
\begin{equation}
\left(\frac{\ell+1}{2},\frac{\ell}{2}\right),
\qquad
\ell=0,1,2,\cdots
\end{equation}
with quantum numbers
\begin{equation}\label{mplus}
m^+_L=-\frac{\ell+1}{2},\cdots,\frac{\ell+1}{2}\qquad m^+_R=-\frac{\ell}{2},\cdots,\frac{\ell}{2}~.
\end{equation}
Also there is another spinor spherical harmonics $Y^-_{\ell;m^-_L,m^-_R}$ on $S^3$ that belong to the representation
\begin{equation}
\left(\frac{\ell}{2},\frac{\ell+1}{2}\right),
\qquad
\ell=0,1,2,\cdots
\end{equation}
with quantum numbers
\begin{equation}\label{mminus}
m^-_L=-\frac{\ell}{2},\cdots,\frac{\ell}{2}\qquad m^-_R=-\frac{\ell+1}{2},\cdots,\frac{\ell+1}{2}~.
\end{equation}
The log partition function of \eqref{partD} can be written as
\begin{equation}\label{propertime}
\ln Z_{\frac{1}{2}}
=
\frac12 \ln \det \mathcal O_{\frac{1}{2}}~,
\qquad
\mathcal O_{\frac{1}{2}}=-\slashed D^{\,2}=-\partial^2_0-\slashed\nabla^2_{S^3}~.
\end{equation}
We will rewrite the determinant in Schwinger proper-time form,
\begin{equation}
\ln Z_{\frac{1}{2}}
=
-\frac12 \int_0^\infty \frac{ds}{s}\,
{\rm Tr}\left(e^{-s{\mathcal O}_{\frac{1}{2}}}\right).
\end{equation}
In this case, we impose twisted anti-periodic boundary conditions along the thermal circle in the following way:
\begin{equation}\label{twistedbc}
\psi(\tau+\beta,\Omega)
=
-e^{\,\beta\,\omega_1 J_1+\beta\,\omega_2 J_2}\,
\psi(\tau,\Omega).
\end{equation}
Here $\Omega$ denotes the angular coordinate  dependence of the wave function on $S^3$.

To compute the partition function, we sum over the eigen spectrum of the kinetic operator. We therefore begin by writing the mode expansion of the Dirac field on $S^1_{\beta}\times S^3$:
\begin{equation}
\psi(\tau,\Omega)
=
\sum_{n\in\mathbb Z}
\left(\sum_{\ell,m^+_L,m^+_R}e^{i\omega^+_n\tau}c^+_{n;\ell;m^+_L,m^+_R}Y^+_{\ell;m^+_L,m^+_R}(\Omega)+\sum_{\ell,m^-_L,m^-_R}e^{i\omega^-_n\tau}c^-_{n;\ell;m^-_L,m^-_R}Y^-_{\ell;m^-_L,m^-_R}(\Omega)\right).
\end{equation}
The anti-periodic boundary conditions implies
\begin{equation}
\omega^{\pm}_n=\frac{(2n+1)\pi}{\beta}-i\left[(\omega_1+\omega_2)m^\pm_L+(\omega_1-\omega_2)m^\pm_R\right]~,\qquad n\in\mathbb{Z}    
\end{equation}
and $(m^\pm_L,m^\pm_R)$ take its value from which spinor representation $\pm$ they belong.\\

The complete set of eigenvalues of ${\mathcal O}_{\frac{1}{2}}$ are given by: \footnote{For a spinor on the unit $S^3$,
\begin{equation}
-\slashed{\nabla}_{S^3}^2
\quad\longrightarrow\quad
\left(\ell+\frac{3}{2}\right)^2.
\end{equation}}
\begin{equation}
\lambda^{\pm}_{n;\ell;m^\pm_L,m^\pm_R}
=
\left(
\frac{(2n+1)\pi}{\beta}
-i\mu^{\pm}(m^\pm_L,m^\pm_R)
\right)^2
+
\left(\ell+\frac{3}{2}\right)^2,
\end{equation}
where we introduced two parameters $\mu^{\pm}$ to be
\begin{equation}\label{muspinordef}
\mu^{\pm}(m^\pm_L,m^\pm_R)=(\omega_1+\omega_2)m^\pm_L+(\omega_1-\omega_2)m^\pm_R.
\end{equation}
Substituting the eigenspectrum into the proper-time representation, we find
\begin{equation}
\begin{split}
\ln Z_{\frac{1}{2}}
&=
-\int_0^\infty \frac{ds}{s}
\Bigg[\sum_{n,\ell,m^+_L,m^+_R}
e^{
-s\left(\ell+\frac{3}{2}\right)^2
-s\left(
\frac{(2n+1)\pi}{\beta}
-i\mu^+
\right)^2}\\&~~~~~~~~~~~~~~+\sum_{n,\ell,m^-_L,m^-_R}
e^{
-s\left(\ell+\frac{3}{2}\right)^2
-s\left(
\frac{(2n+1)\pi}{\beta}
-i\mu^-
\right)^2}\Bigg]~.    
\end{split}
\end{equation}
In the above equation, we have multiplied overall a $2$ factor because we have considered Dirac fermion.

Following the same steps from \eqref{poisson} to \eqref{logzsccas},
we find the logarithm of the partition function of the massless Dirac field to be
\begin{equation}
\begin{split}
\ln Z_{\frac{1}{2}}
&=\sum_{k\in\mathbb{Z}\setminus\{0\}}\sum_{\ell}\frac{(-1)^{k+1}}{|k|} e^{-\beta |k|\left(\ell+\frac{3}{2}\right)}\Bigg[\sum_{m^+_L,m^+_R}e^{k\beta\mu^+}+\sum_{m^-_L,m^-_R}e^{k\beta\mu^-}\Bigg]\\&=2\sum_{k=1}^{\infty}\sum_{\ell=0}^{\infty}\frac{(-1)^{k+1}}{k} e^{-\beta k\left(\ell+\frac{3}{2}\right)}\Bigg[\sum_{m^+_L,m^+_R}e^{k\beta\mu^+}+\sum_{m^-_L,m^-_R}e^{k\beta\mu^-}\Bigg]  
\end{split}    
\end{equation}
where we have used $\mu^{\pm}(-m_L,-m_R)=-\mu^{\pm}(m_L,m_R)$ in the second line. Now we can express $(m_L,m_R)$ sums as the product of two $SU(2)$ characters. So we have
\begin{equation}
\sum_{m^+_L,m^+_R}e^{k\beta\mu^+(m_L^+,m_R^+)}=\frac{
\sinh\!\left(\frac{k\beta(\ell+2)(\omega_1+\omega_2)}{2}\right)
}{
\sinh\!\left(\frac{k\beta(\omega_1+\omega_2)}{2}\right)
}
\frac{
\sinh\!\left(\frac{k\beta(\ell+1)(\omega_1-\omega_2)}{2}\right)
}{
\sinh\!\left(\frac{k\beta(\omega_1-\omega_2)}{2}\right)
}
\end{equation}
and similarly we also have
\begin{equation}
\sum_{m^-_L,m^-_R}e^{k\beta\mu^-(m^-_L,m^-_R)}=\frac{
\sinh\!\left(\frac{k\beta(\ell+1)(\omega_1+\omega_2)}{2}\right)
}{
\sinh\!\left(\frac{k\beta(\omega_1+\omega_2)}{2}\right)
}
\frac{
\sinh\!\left(\frac{k\beta(\ell+2)(\omega_1-\omega_2)}{2}\right)
}{
\sinh\!\left(\frac{k\beta(\omega_1-\omega_2)}{2}\right)
}~.
\end{equation}
Performing the $\ell$ sum, the full partition function is given by
\begin{equation}
\begin{split}
\ln Z_{\frac{1}{2}}&=\sum_{k=1}^{\infty}\frac{(-1)^{k+1}}{k}\frac{8e^{-3k\beta/2}\left(1-e^{-k\beta}\right)\cosh\left(\frac{k\beta\omega_1}{2}\right)\cosh\left(\frac{k\beta\omega_2}{2}\right)}{\left(1-e^{-k\beta(1-\omega_1)}\right)\left(1-e^{-k\beta(1+\omega_1)}\right)
 \left(1-e^{-k\beta(1-\omega_2)}\right)\left(1-e^{-k\beta(1+\omega_2)}\right)}\\&=\sum_{k=1}^{\infty}\frac{(-1)^{k+1}}{k}z_{\frac{1}{2}}(k)  
\end{split}   
\end{equation}
where $z_{\frac{1}{2}}(k)$ is the single particle partition function as given by
\begin{equation}\label{zspd}
 z_{\frac{1}{2}}(k)=\frac{8e^{-3k\beta/2}\left(1-e^{-k\beta}\right)\cosh\left(\frac{k\beta\omega_1}{2}\right)\cosh\left(\frac{k\beta\omega_2}{2}\right)}{\left(1-e^{-k\beta(1-\omega_1)}\right)\left(1-e^{-k\beta(1+\omega_1)}\right)
 \left(1-e^{-k\beta(1-\omega_2)}\right)\left(1-e^{-k\beta(1+\omega_2)}\right)}~.   
\end{equation}

The single-particle partition function can be very easily understood using the state-operator map. The conformal weight of the massless Dirac field is $\Delta_{\psi}=\frac{3}{2}$. The four-component Dirac spinor carries $SO(4)\simeq SU(2)_L\times SU(2)_R$ the angular weight $\left(\pm \frac{1}{2},\pm \frac{1}{2}\right).$ Let us define $x_i=e^{k\beta\omega_i}$, for $i=1,2$.

The spin character is given by
\begin{align}\label{spinch}
\chi_{\psi}(x_1,x_2)
&=
\sum_{\sigma_1,\sigma_2=\pm 1}
x_1^{\frac{\sigma_1}{2}}\,
x_2^{\frac{\sigma_2}{2}}\nonumber\\
&=
\left(x_1^{1/2}+x_1^{-1/2}\right)
\left(x_2^{1/2}+x_2^{-1/2}\right)\nonumber\\
&=4\cosh\left(\frac{k\beta\omega_1}{2}\right)\cosh\left(\frac{k\beta\omega_2}{2}\right)
\end{align}
We now act with the derivatives, and the four derivative components have weights 
\begin{align}
    q\,x_1\qquad q\,x^{-1}_1 \qquad q\,x_2\qquad q\,x^{-1}_2
\end{align}
which have corresponding angular charges  $(\pm 1,0)$ and $(0,\pm 1)$.
Massless Dirac fermions also satisfy the equation of motion 
\begin{align}
    \gamma^{\mu}D_{\mu}\psi=0.
\end{align}
Therefore, we will also have to remove the null descendants of weight $\Delta_{\psi}=\frac{5}{2}$. Therefore, the single particle partition function of the Dirac field is given by
\begin{align}
    z_{\frac{1}{2}}(k)&=\frac{2q^{\frac{3}{2}}\left(1-q\right)\times \chi_{\psi}}{\left(1-q\, x_1\right)\left(1-q\, x^{-1}_1\right)\left(1-q\, x_2\right)\left(1-q\, x^{-1}_2\right)},
\end{align}
which reproduces \eqref{zspd}. We have multiplied by an extra $2$ factor because each component of the Dirac fermion is complex, which is equivalent to two real degrees of freedom.


We now study the partition function in the semi-universal limit \eqref{limit}.
In this limit, the leading divergent part of the partition function is
\begin{equation}
\ln Z_{\frac{1}{2}}
=
\frac{\beta^2 h_{\frac{1}{2}}(\beta)}{\nu_1\nu_2}
+
O\!\left(\frac{1}{\nu_i}\right)
+\cdots ,
\end{equation}
where, the residue function is given by
\begin{equation}\label{hbetasc}
h_{\frac{1}{2}}(\beta)
=
\frac{1}{\beta^2}
\sum_{k=1}^{\infty}
\frac{2(-1)^{k+1}e^{-k\beta/2}}
{k^3(1-e^{-k\beta})}
=
\frac{1}{\beta^2}
\sum_{k=1}^{\infty}
\frac{(-1)^{k+1}}{k^3\sinh\left(\frac{k\beta}{2}\right)} .
\end{equation}

We note that the residue function is a well behaved function of $\beta$. The large $k$-terms in the sum do not grow exponentially. This phenomenon can be attributed to the absence of the negative twist states,in the theory. We discuss this in detail in section \ref{ansc}.


\subsection{Conformal higher derivative fermions}
The Weyl-invariant three-derivative fermion was originally encountered in the context of extended conformal supergravity \cite{Bergshoeff:1980is}, and was later studied in four dimensions for Majorana fermions in \cite{Fradkin:1981jc}. According to the classification introduced in the introduction, this theory falls into \textit{Class 4}: it violates the naive unitarity bound and, at the same time, contains negative-twist states.

Let us study the partition function of $3$-derivative conformal fermion on $S^1_{\beta}\times S^3$ in the `semi-universal' limit \eqref{limit}. We also study the properties of the residue function of the leading divergent term of the partition function.

The action of the conformal fermion with a cubic-derivative kinetic term in $d=4$-dimensions is given by \cite{Fradkin:1981jc}
\begin{align}
S_{\Psi}
&=
i\int d^4x\,\sqrt{g}\,\bar\psi\,\mathcal O_{\Psi}\psi ,
\\
\mathcal O_{\Psi}
&=
\slashed D^{\,3}
+
\left(
R^{\mu\nu}
-\frac16 R g^{\mu\nu}
\right)\gamma_\mu D_\nu~.
\end{align}
On $S^1_\beta\times S^3$, the kinetic operator  is expressed as
\begin{align}
\mathcal O_{\Psi}
&=\slashed D^{\,3}
-\mathbf{\slashed D}
+2{\slashed \nabla}_{S^3}~.
\end{align}
Using the following identity
\begin{align}
\left\{\mathbf{\slashed \nabla}_{S^3},\gamma^0\right\}=0,
\end{align}
we find that the determinant of this operator factorizes as
\begin{equation}
\begin{split}
Z_{\rm \Psi}=\det\big[i\mathcal O_{\Psi}\big]
&=\left[\det\left(-\mathcal{O}_{\rm \Psi}^2\right)\right]^{1/2}~.    
\end{split}
\end{equation}
We can also decompose $\mathcal{O}_{\rm \Psi}^2$ as the following
\begin{equation}
\mathcal{O}_{\rm \Psi}^2=\left[\partial_0^2+\left(\slashed{\nabla}_{S^3}+i\right)^2\right]\left[\partial_0^2+\slashed{\nabla}^2_{S^3}\right] \left[\partial_0^2+\left(\slashed{\nabla}_{S^3}-i\right)^2\right]~.    
\end{equation}
Hence, we can write the fermionic determinant as
\begin{equation}\label{hdfer}
\det\big[i\mathcal O_{\Psi}\big]=\left\{\det\left(-\left[\partial_0^2+\left(\slashed{\nabla}_{S^3}+i\right)^2\right]\right)\det\left(-\left[\partial_0^2+\slashed{\nabla}_{S^3}^2\right]\right)\det\left(-\left[\partial_0^2+\left(\slashed{\nabla}_{S^3}-i\right)^2\right]\right)\right\}^{1/2}    
\end{equation}
Using \eqref{hdfer}, we can express the partition function of the cubic derivative fermionic fields on $S^1_{\beta}\times S^3$ in the following way:
\begin{equation}\label{higherderivativefer}
\begin{split}
\ln Z_{\rm \Psi}&=\frac{1}{2}\ln \det\left(-\left[\partial_0^2+\slashed{\nabla}_{S^3}^2\right]\right)+\frac{1}{2}\ln \det\left(-\left[\partial_0^2+\left(\slashed{\nabla}_{S^3}-i\right)^2\right]\right)\\&~~~~~~~~~~~~~~~~~~~~~~~+\frac{1}{2}\ln \det\left(-\left[\partial_0^2+\left(\slashed{\nabla}_{S^3}+i\right)^2\right]\right)~.     
\end{split}  
\end{equation}
We have already computed the first term of the partition function in the massless Dirac field. Let us now the second term written in terms of the Schwinger parameter, as given by
\begin{equation}
\frac12 \ln \det \mathcal{O}_1=-\frac12 \int_0^\infty \frac{ds}{s}\,
{\rm Tr}\left(e^{-s\mathcal {O}_1}\right)~,
\qquad
\mathcal{O}_1=-\partial^2_0-\left(\slashed\nabla_{S^3}-i\right)^2~.    
\end{equation}
Imposing the twisted anti-periodic boundary condition, we can write down the eigenvalues of $\mathcal{O}_1$ as given by
\begin{equation}
\begin{split}
\lambda^{+}_{n;\ell;m_L,m_R}
&=
\left(
\frac{(2n+1)\pi}{\beta}
-i\mu^{+}(m_L,m_R)
\right)^2
+
\left(\ell+\frac{5}{2}\right)^2~,\\\lambda^{-}_{n;\ell;m_L,m_R}
&=
\left(
\frac{(2n+1)\pi}{\beta}
-i\mu^{-}(m_L,m_R)
\right)^2
+
\left(\ell+\frac{1}{2}\right)^2~.    
\end{split}
\end{equation}
where, we have defined $\mu^{\pm}$ in the following way:
\begin{equation}
\mu^\pm(m_L,m_R)
=
(\omega_1+\omega_2)m^\pm_L
+
(\omega_1-\omega_2)m^\pm_R.
\end{equation}
The range of  $m_{L/R}^{\pm}$ are given in \eqref{mplus} and \eqref{mminus}.

Substituting the eigenspectrum into the above representation in the Schwinger proper-time form of the determinant, we find
\begin{equation}
\begin{split}
\frac12 \ln \det \mathcal{O}_1
&=
-\int_0^\infty \frac{ds}{s}
\Bigg[\sum_{n,\ell,m_L,m_R}
e^{
-s\left(\ell+\frac{5}{2}\right)^2
-s\left(
\frac{(2n+1)\pi}{\beta}
-i\mu^+
\right)^2}\\&~~~~~~~~~~~~~~+\sum_{n,\ell,m_L,m_R}
e^{
-s\left(\ell+\frac{1}{2}\right)^2
-s\left(
\frac{(2n+1)\pi}{\beta}
-i\mu^-
\right)^2}\Bigg]~.    
\end{split}
\end{equation}
Similarly, we can express the third term in \eqref{higherderivativefer} as
\begin{equation}
\begin{split}
\frac12 \ln \det \mathcal{O}_2
&=
-\int_0^\infty \frac{ds}{s}
\Bigg[\sum_{n,\ell,m_L,m_R}
e^{
-s\left(\ell+\frac{1}{2}\right)^2
-s\left(
\frac{(2n+1)\pi}{\beta}
-i\mu^+
\right)^2}\\&~~~~~~~~~~~~~~+\sum_{n,\ell,m_L,m_R}
e^{
-s\left(\ell+\frac{5}{2}\right)^2
-s\left(
\frac{(2n+1)\pi}{\beta}
-i\mu^-
\right)^2}\Bigg]     
\end{split}    
\end{equation}
where $\mathcal{O}_2=-\partial^2_0-\left(\slashed\nabla_{S^3}+i\right)^2$. Since we have considered Dirac fermion, we have multiplied factor $2$ in the R.H.S of the above two equations.

Following the same steps from \eqref{poisson} to \eqref{logzsccas}, and performing manipulations with the $SU(2)$ characters in a similar fashion as done in the previous section, and performing the $\ell$ sum, the above expression is given by
\begin{equation}
\begin{split}
&\frac12 \ln \det \mathcal{O}_1+\frac12 \ln \det \mathcal{O}_2\\&=\sum_{k=1}^{\infty}\frac{(-1)^{k+1}}{k}\frac{8\left(e^{-k\beta/2}+e^{-5k\beta/2}\right)\left(1-e^{-k\beta}\right)\cosh\left(\frac{k\beta\omega_1}{2}\right)\cosh\left(\frac{k\beta\omega_2}{2}\right)}{\left(1-e^{-k\beta(1-\omega_1)}\right)\left(1-e^{-k\beta(1+\omega_1)}\right)
 \left(1-e^{-k\beta(1-\omega_2)}\right)\left(1-e^{-k\beta(1+\omega_2)}\right)}~. 
\end{split}   
\end{equation}
So the full log partition function for the Conformal higher derivative fermions is given by
\begin{equation}
\begin{split}
\ln Z_{\rm \Psi}&=\sum_{k=1}^{\infty}\frac{(-1)^{k+1}}{k}\frac{8\left(e^{-k\beta/2}+e^{-3k\beta/2}+e^{-5k\beta/2}\right)\left(1-e^{-k\beta}\right)\cosh\left(\frac{k\beta\omega_1}{2}\right)\cosh\left(\frac{k\beta\omega_2}{2}\right)}{\left(1-e^{-k\beta(1-\omega_1)}\right)\left(1-e^{-k\beta(1+\omega_1)}\right)
 \left(1-e^{-k\beta(1-\omega_2)}\right)\left(1-e^{-k\beta(1+\omega_2)}\right)}\\&=\sum_{k=1}^{\infty}\frac{(-1)^{k+1}}{k}\frac{8e^{-k\beta/2}\left(1-e^{-3k\beta}\right)\cosh\left(\frac{k\beta\omega_1}{2}\right)\cosh\left(\frac{k\beta\omega_2}{2}\right)}{\left(1-e^{-k\beta(1-\omega_1)}\right)\left(1-e^{-k\beta(1+\omega_1)}\right)
 \left(1-e^{-k\beta(1-\omega_2)}\right)\left(1-e^{-k\beta(1+\omega_2)}\right)}\\&=\sum_{k=1}^{\infty}\frac{(-1)^{k+1}}{k}z_{\rm \Psi}(k)     
\end{split}   
\end{equation}
where $z_{\rm \Psi}(k)$ is the single particle partition function as given by
\begin{equation}
 z_{\rm \Psi}(k)=\frac{8e^{-k\beta/2}\left(1-e^{-3k\beta}\right)\cosh\left(\frac{k\beta\omega_1}{2}\right)\cosh\left(\frac{k\beta\omega_2}{2}\right)}{\left(1-e^{-k\beta(1-\omega_1)}\right)\left(1-e^{-k\beta(1+\omega_1)}\right)
 \left(1-e^{-k\beta(1-\omega_2)}\right)\left(1-e^{-k\beta(1+\omega_2)}\right)}~.   
\end{equation}

The single-particle partition function of the cubic derivative fermion can be understood very easily using the operator counting method. From the action, we can see that the cubic derivative fermionic field has a conformal dimension $\Delta_{\psi}=\frac{1}{2}$. Acting with the derivatives gives us the unrestricted descendant module
\begin{equation}
\frac{1}
{(1-qx_1)(1-qx_{1}^{-1})
 (1-qx_2)(1-qx^{-1}_2)}.
\end{equation}
Here $q=e^{-k\beta}$, $x_i=e^{\beta\omega_i}$.
For the cubic derivative fermion , the equation of motion is
\begin{equation}
\slashed {D}^3 \psi=0.
\end{equation}
Therefore, we have to remove the null states corresponding to the equation of motion, which comes with the same denominator but with the numerator $q^{\Delta_{\Psi}+3}$. The single-particle partition function of the cubic derivative fermion can be expressed as
\begin{align}
    z_{\Psi}(k)&=\frac{2q^{\Delta_{\Psi}}\left(1-q^3\right)\times \chi_{\psi}}{(1-qx_1)(1-qx_{1}^{-1})
 (1-qx_2)(1-qx^{-1}_2)},
\end{align}
where $\chi_{\psi}$ corresponds to the spin character given in \eqref{spinch}. We note that, the partition function obtained using the operator counting method is the same as the partition function obtained using the mode-sum approach.


We now study the partition function in the semi-universal limit \eqref{limit}.
In this limit, the leading divergent part of the partition function is
\begin{equation}
\ln Z_{\rm \Psi}
=
\frac{\beta^2 h_{\rm \Psi}(\beta)}{\nu_1\nu_2}
+
O\!\left(\frac{1}{\nu_i}\right)
+\cdots ,
\end{equation}
where, the residue function of the leading singularity in the semi-universal limit of the partition function is given by
\begin{equation}\label{hbetasc}
h_{\rm \Psi}(\beta)
=
\frac{2}{\beta^2}
\sum_{k=1}^{\infty}
\frac{(-1)^{k+1}\left(e^{k\beta/2}+e^{-k\beta/2}+e^{-3k\beta/2}\right)}
{k^3(1-e^{-k\beta})}.
\end{equation}

Note that the residue function is not well behaved. The large $k$-terms of the sum grow as $(-1)^{k+1}\frac{e^{\frac{k\beta}{2}}}{k^3}$ (with alternating signs). This indicates the presence of a negative twist state  with twist $\tau=-\frac{1}{2}$  in the single-particle partition function.  We discuss this more in section \ref{ansc}.


\section{Semi-universality of a Weyl invariant vector field}\label{vc}
In this section, we study the partition function of conformal vector fields, namely the free Maxwell field on $S^1_{\beta}\times S^3$ and the conformal higher derivative vector field on $S^1_{\beta}\times S^5$ in the semi-universal limit \eqref{limit}, using both the mode-sum and the operator counting approaches. We extract the residue function of the leading poles of the partition function in this limit.
\subsection{Maxwell field on $S^1_{\beta}\times S^3$ }
 According to the classification introduced in the introduction, this theory falls into \textit{Class 1}: it obeys the naive unitarity bound and, at the same time, does not contain negative-twist states.
The action of the free Maxwell's theory in $d=4$ dimension is given by
\begin{align}
S_{1}
&=
-\frac{1}{4}\int d^4x\,\sqrt{g}\,F_{\mu\nu}F^{\mu\nu} ,
\\
F_{\mu\nu}
&=D_\mu A_{\nu}-D_\nu A_{\mu}~.
\end{align}
The gauge fixed partition function in the temporal gauge $A_{\tau}=0$, is given by \cite{Beccaria:2014jxa, David:2020mls}
\begin{equation}
Z_{1}=\frac{1}{\left[\det\mathcal{O}^{\perp}_1\right]^{1/2}}    
\end{equation}
where $\mathcal{O}^{\perp}_1$ is the kinetic operator corresponding to the transverse spin-1 Laplacian and is given by
\begin{equation}
 \mathcal{O}^{\perp}_{1}=\big(-\partial_0^2-\nabla^2_{S^3}+2\big)   
\end{equation}
We will now evaluate the partition function using the mode-sum approach.  Transverse vector harmonics $Y^{i;+}_{\ell;m^+_L,m^+_R}$ on $S^3$ belong to the representation
\begin{equation}
\left(\frac{\ell+2}{2},\frac{\ell}{2}\right)~,
\qquad
\ell=0,1,2,\cdots
\end{equation}
with quantum numbers
\begin{equation}
m^+_L=-\frac{\ell+2}{2},\cdots,\frac{\ell+2}{2}\qquad m^+_R=-\frac{\ell}{2},\cdots,\frac{\ell}{2}~.
\end{equation}
Also there is another transverse vector spherical harmonics $Y^{i;-}_{\ell;m^-_L,m^-_R}$ on $S^3$ that belong to the representation
\begin{equation}
\left(\frac{\ell}{2},\frac{\ell+2}{2}\right)~,
\qquad
\ell=0,1,2,\cdots
\end{equation}
with quantum numbers
\begin{equation}
m^-_L=-\frac{\ell}{2},\cdots,\frac{\ell}{2}\qquad m^-_R=-\frac{\ell+2}{2},\cdots,\frac{\ell+2}{2}~.
\end{equation}
We impose twisted periodic boundary conditions along the thermal circle in the following way:
\begin{equation}\label{twistedbcvector}
A^i(\tau+\beta,\Omega)
=
e^{\,\beta\,\omega_1 J_1+\beta\,\omega_2 J_2}\,
A^i(\tau,\Omega).
\end{equation}
To compute the partition function, we sum over the eigenspectrum of the kinetic operator. We therefore begin by writing the mode expansion of the scalar field on $S^1_{\beta}\times S^3$:
\begin{equation}
A^i(\tau,\Omega)
=
\sum_{n\in\mathbb Z}
\left(\sum_{\ell,m^+_L,m^+_R}e^{i\omega^+_n\tau}c^+_{n;\ell;m^+_L,m^+_R}Y^{i;+}_{\ell;m^+_L,m^+_R}(\Omega)+\sum_{\ell,m^-_L,m^-_R}e^{i\omega^-_n\tau}c^-_{n;\ell;m^-_L,m^-_R}Y^{i;-}_{\ell;m^-_L,m^-_R}(\Omega)\right).
\end{equation}
The periodic boundary conditions implies
\begin{equation}
\omega^{\pm}_n=\frac{2n\pi}{\beta}-i\left[(\omega_1+\omega_2)m^\pm_L+(\omega_1-\omega_2)m^\pm_R\right]~,\qquad n\in\mathbb{Z}    
\end{equation}
and $(m^\pm_L,m^\pm_R)$ take its value from which transverse Vector representation $\pm$ they belong.

The complete set of eigenvalues of $\mathcal O^{\perp}_1$ are given by: \footnote{For a Vector on the unit $S^3$,
\begin{equation}
-\nabla_{S^3}^2+2
\quad\longrightarrow\quad
\left(\ell+2\right)^2.
\end{equation}}
\begin{equation}
\lambda^{\pm}_{n;\ell;m^\pm_L,m^\pm_R}
=
\left(
\frac{2n\pi}{\beta}
-i\mu^{\pm}(m^\pm_L,m^\pm_R)
\right)^2
+
\left(\ell+2\right)^2,
\end{equation}
where we introduced two parameters $\mu^{\pm}$ to be
\begin{equation}\label{muspinordef}
\mu^{\pm}(m^\pm_L,m^\pm_R)=(\omega_1+\omega_2)m^\pm_L+(\omega_1-\omega_2)m^\pm_R.
\end{equation}
Now performing the similar method as done in the subsection (\ref{dirac}), we obtain
\begin{equation}\label{partitionfunctionvector}
\ln Z_{1}=\sum_{k=1}^{\infty}\sum_{\ell=0}^{\infty}\frac{1}{k} e^{-\beta k\left(\ell+2\right)}\Bigg[\sum_{m^+_L,m^+_R}e^{k\beta\mu^+}+\sum_{m^-_L,m^-_R}e^{k\beta\mu^-}\Bigg]~.    
\end{equation}
Here also, we can express $(m^{\pm}_L,m^{\pm}_R)$ sums as the product of two $SU(2)$ characters. So we have
\begin{equation}
\sum_{m^+_L,m^+_R}e^{k\beta\mu^+(m_L^+,m_R^+)}=\frac{
\sinh\!\left(\frac{k\beta(\ell+3)(\omega_1+\omega_2)}{2}\right)
}{
\sinh\!\left(\frac{k\beta(\omega_1+\omega_2)}{2}\right)
}
\frac{
\sinh\!\left(\frac{k\beta(\ell+1)(\omega_1-\omega_2)}{2}\right)
}{
\sinh\!\left(\frac{k\beta(\omega_1-\omega_2)}{2}\right)
}
\end{equation}
and similarly we have
\begin{equation}
\sum_{m^-_L,m^-_R}e^{k\beta\mu^-(m^-_L,m^-_R)}=\frac{
\sinh\!\left(\frac{k\beta(\ell+1)(\omega_1+\omega_2)}{2}\right)
}{
\sinh\!\left(\frac{k\beta(\omega_1+\omega_2)}{2}\right)
}
\frac{
\sinh\!\left(\frac{k\beta(\ell+3)(\omega_1-\omega_2)}{2}\right)
}{
\sinh\!\left(\frac{k\beta(\omega_1-\omega_2)}{2}\right)
}~.
\end{equation}
Performing the $\ell$ sum, the full partition function is given by
\begin{equation}
\begin{split}
\ln Z_{1}&=\sum_{k=1}^{\infty}\frac{1}{k}\frac{2e^{-2k\beta}\Big[1+e^{-2k\beta}+2\cosh(k\beta\omega_1)\cosh(k\beta\omega_2)-2e^{-k\beta}\big(\cosh(k\beta\omega_1)+\cosh(k\beta\omega_2)\big)\Big]}{\left(1-e^{-k\beta(1-\omega_1)}\right)\left(1-e^{-k\beta(1+\omega_1)}\right)
 \left(1-e^{-k\beta(1-\omega_2)}\right)\left(1-e^{-k\beta(1+\omega_2)}\right)}\\&=\sum_{k=1}^{\infty}\frac{1}{k}z_{1}(k)  
\end{split}   
\end{equation}
where $z_{1}(k)$ is the single particle partition function as given by
\begin{align}\label{maxpartmode}
 z_{1}(k)&=\frac{2e^{-2k\beta}\Big[1+e^{-2k\beta}+2\cosh(k\beta\omega_1)\cosh(k\beta\omega_2)-2e^{-k\beta}\big(\cosh(k\beta\omega_1)+\cosh(k\beta\omega_2)\big)\Big]}{\left(1-e^{-k\beta(1-\omega_1)}\right)\left(1-e^{-k\beta(1+\omega_1)}\right)
 \left(1-e^{-k\beta(1-\omega_2)}\right)\left(1-e^{-k\beta(1+\omega_2)}\right)}\nonumber\\
 &=\frac{
q^2\left[
(x_1+x_1^{-1})(x_2+x_2^{-1})+2
\right]
-
2q^3\left[
x_1+x_1^{-1}+x_2+x_2^{-1}
\right]+2q^4
}
{
(1-qx_1)(1-qx_1^{-1})(1-qx_2)(1-qx_2^{-1})
},
\end{align}
where $q=e^{-k\beta}$ and $x_i=e^{k\beta\omega_i}$ and we note that, the first term with $q^2$ in the numerator is nothing but the sum of two spin-1 $SU(2)$ character whereas the second term with $q^3$ is the product of two spin-$\frac{1}{2}$ $SU(2)$-character.

The partition function of the free Maxwell field can also be obtained using the operator counting method. The gauge invariant primary operator is the field strength with the scaling dimension $\Delta_F=2$, and under $SO(4)\simeq SU(2)_L\times SU(2)_R$, the field strength decomposes into the self-dual and the anti self-dual pieces, which transform as $(1,0)\oplus(0,1)$. Therefore, the spin character of the Maxwell primary is
\begin{align}
    \chi_F(x_L,x_R)  =  \chi_1(x_L)+\chi_1(x_R),
\end{align}
where $\chi_1(x_L/x_R)$ are the spin-1 $SU(2)$ character with $x_L=e^{k\beta\frac{(\omega_1+\omega_2)}{2}}$ and $x_R=e^{k\beta\frac{(\omega_1-\omega_2)}{2}}$.
The unrestricted descendant module is obtained by acting with arbitrary derivatives on the primary. Since, the derivative transforms as $\left(\frac{1}{2},\frac{1}{2}\right)$, and each derivative contributes to either of the four weights,i.e, $qx_1$, $qx^{-1}_1$, $q x_{2}$ or $qx^{-1}_2$, the unrestricted descendant module gives 

$$\frac{1}{\left(1-q\, x_1\right)\left(1-q\, x^{-1}_1\right)\left(1-q\, x_2\right)\left(1-q\, x^{-1}_2\right)}~.$$
We should also remove the null states corresponding to the equation of motion and the Bianchi identity
\begin{equation}\label{eommax}
    \begin{split}
        D_{\mu}F^{\mu\nu}&=0~,\\
        D_{\mu}\tilde{F}^{\mu\nu}&=0~.
    \end{split}
\end{equation}
Note that these are level-one descendants and transform as a vector which is $\left(\frac{1}{2},\frac{1}{2}\right)$ and comes with the weight $q^{\Delta_F+1}$. Therefore, the spin character associated with them will be $\chi_{\frac{1}{2}}(x_L)\,\chi_{\frac{1}{2}}(x_R)$.
Therefore, the single particle partition function of the free Maxwell field, after removing null states, is given by:
\begin{align}\label{maxwellopcount}
z_1(q,x_1,x_2)&=
\frac{
q^2\left[\chi_1(x_L)+\chi_1(x_R)\right]-2q^3\chi_{\frac{1}{2}}(x_L)\chi_{\frac{1}{2}}(x_R)+2q^4}
{(1-qx_1)(1-qx_1^{-1})(1-qx_2)(1-qx_2^{-1})}.
\end{align}
In the numerator, we have further subtracted the null-states corresponding to the following equations
\begin{align}
   D_\mu D_\nu F^{\mu\nu}=0, \qquad
D_\mu D_\nu \widetilde F^{\mu\nu}=0.
\end{align}
Note that, the partition function obtained using the operator counting method agrees with the mode-sum approach in \eqref{maxpartmode}.


Let us now study the partition function of the Maxwell field in the semi-universal limit \eqref{limit}.
\begin{equation}
\ln Z_{1}=\frac{\beta^2 h_1(\beta)}{\nu_1\nu_2}+O\left(\frac{1}{\nu_i}\right)    
\end{equation}
where the residue function $h_1(\beta)$ is given by
\begin{equation}
h_1(\beta)=\frac{1}{\beta^2}\sum_{k=1}^{\infty}\frac{\coth(k\beta)}{k^3}~. 
\end{equation}

We observe that the residue function $h_1(\beta)$ is the same as the residue function obtained from the 4-derivative conformal scalar field in \eqref{h4derphi}. Similar to the 4-derivative scalar, the Maxwell theory also does not have any negative twist state in the single-particle sector; therefore, the residue function is also well behaved. The partition function is also well behaved in every sense.

\subsection{Conformal higher derivative vector field on $S^1_{\beta}\times S^5$}
 According to the classification introduced in the introduction, this theory falls into \textit{Class 2}: it violates the naive unitarity bound, but it does not contain any negative-twist states.
The Euclidean action of a Weyl invariant higher derivative vector field in $d=6$-dimensions is given by \cite{Beccaria:2015uta, Beccaria:2017dmw}
\begin{equation}
\begin{split}
&S_1=\int d^6x\sqrt{g}\left[D_\lambda F^{\lambda\mu}D^{\nu}F_{\nu\mu}-\left(R_{\mu\nu}-\frac{1}{5}g_{\mu\nu}R\right)F^{\mu\lambda}F^\nu_{~\lambda}\right]\\&F_{\mu\nu}
=D_\mu A_{\nu}-D_\nu A_{\mu}~.      
\end{split}  
\end{equation}
Note that the kinetic term of the action has four derivatives acting on the gauge field.
The gauge fixed partition function on $S^1\times S^5$ is given by \cite{Beccaria:2017dmw}
\begin{equation}
Z_{1}=\frac{1}{\left[\det\mathcal{O}^{\perp}_1\right]^{1/2}}    
\end{equation}
where the kinetic operator $\mathcal{O}^{\perp}$ is given by
\begin{equation}
 \begin{split}
 \mathcal{O}^{\perp}_1&=\left(-\partial_0^2-\nabla^2_{S^5}+4\right)^2-4\partial_0^2\\&=\left(-\partial_0^2-2\partial_0-\nabla^2_{S^5}+4\right)\left(-\partial_0^2+2\partial_0-\nabla^2_{S^5}+4\right)\\&=\Big[-\big(\partial_0-1\big)^2-\nabla^2_{S^5}+5\Big]\Big[-\big(\partial_0+1\big)^2-\nabla^2_{S^5}+5\Big]~.      
 \end{split} 
\end{equation}
We can express log of the partition function as
\begin{equation}\label{higherderivative}
\begin{split}
\ln Z_{1}&=-\frac{1}{2}\ln \det\Big[-\big(\partial_0-1\big)^2-\nabla^2_{S^5}+5\Big]-\frac{1}{2}\ln \det\Big[-\big(\partial_0+1\big)^2-\nabla^2_{S^5}+5\Big]~.     
\end{split}  
\end{equation}
To incorporate the angular chemical potentials $\omega_1,\,\omega_2$ and $\omega_3$ associated with the three commuting rotations of $S^5$, here also we impose twisted periodic boundary conditions along the thermal circle in the following way:
\begin{equation}\label{twistedbcvector1}
A^i(\tau+\beta,\Omega)
=
e^{\,\beta\,\omega_1 J_1+\beta\,\omega_2 J_2+\beta\,\omega_3 J_3}\,
A^i(\tau,\Omega)~.
\end{equation}
Here $\Omega$ denotes the angular coordinate on $S^5$, and $J_1,\,J_2$ and $J_3$ are the generators of rotations in three orthogonal planes of $\mathbb{R}^6$, in which $S^5$ is embedded. It gives us a convenient choice of Cartan generators of the isometry group $SO(6)$. We know that the transverse vector representation of $SO(6)$ has highest weight $(\ell+1,1,0)$.\\ The log of the partition function is given by
\begin{equation}
\ln Z_1=-\frac{1}{2}\ln\det\mathcal{O}_1-\frac{1}{2}\ln\det\mathcal{O}_2    
\end{equation}
where $\mathcal{O}_1=-(\partial_0+1)^2-\nabla_{S^5}^2+5$ and $\mathcal{O}_2=-(\partial_0-1)^2-\nabla_{S^5}^2+5$. Following the same methods as described in the previous sections, we have evaluated the partition function in \ref{app1}. We present the final form of the partition function below:
\begin{equation}
\begin{split}
\ln Z_1=\sum_{k=1}^{\infty}\frac{1}{k}z_1(k)    
\end{split}    
\end{equation}
where the single particle partition function for the higher derivative vector is 
\begin{equation}
\begin{split}
&z_1(k)\\&=\frac{e^{-2k\beta}\left(1+e^{-2k\beta}\right)}{\prod_{i=1}^{3}\left(1-e^{-k\beta(1-\omega_i)}\right)\left(1-e^{-k\beta(1+\omega_i)}\right)
 }\\&\times\Big[3+y_1(k)y_2(k)+y_2(k)y_3(k)+y_3(k)y_1(k)-e^{-k\beta}\big(y_1(k)y_2(k)y_3(k)+3y_1(k)+3y_2(k)+3y_3(k)\big)\\&~~~~~+e^{-2k\beta}\big(4+y_1(k)y_2(k)+y_2(k)y_3(k)+y_3(k)y_1(k)\big)-e^{-3k\beta}\big(y_1(k)+y_2(k)+y_3(k)\big)+e^{-4k\beta}\Big]    
\end{split}    
\end{equation}
where $y_i(k)$ is defined as
\begin{equation}
y_i(k)\equiv 2\cosh(k\beta\omega_i)~,\qquad i=1,2,3~.    
\end{equation}
We will study $\ln Z_1$ in the semi-universal limit. In $d=6$, this limit is defined as \cite{Anand:2025mfh}
\begin{equation}\label{limitvec}
    \nu_i\rightarrow 0~, \qquad \text{where} \qquad \nu_i=\beta(1-\omega_i)~,\qquad \frac{\nu_1}{\nu_2},~\frac{\nu_2}{\nu_3},~\frac{\nu_3}{\nu_1}\qquad\text{fixed}~.
\end{equation}
In this limit, $\ln Z_1$ behaves like
\begin{equation}
\ln Z_1=\frac{\beta^2h^{d=6}_1(\beta)}{\nu_1\nu_2\nu_3}+O\left(\frac{1}{\nu_i\nu_j}\right)    
\end{equation}
where the function $h^{d=6}_1(\beta)$ is given by
\begin{equation}
h^{d=6}_1(\beta)=\frac{1}{\beta^2}\sum_{k=1}^{\infty}\frac{2\coth^{2}(k\beta)}{k^4}~.    
\end{equation}
We note that the residue function $h^{d=6}_1(\beta)$ of the conformal higher derivative vector field does not exhibit exponential growth in large $k$, and therefore,is well behaved in the semi-universal limit. Although the theory violates the naive unitarity bound, it does not contain any negative twist state and therefore does not exhibit pathological behavior in the semi-universal limit.

\section{Semi-universality of Weyl gravitino on $S^1_\beta \times S^3$}\label{grt}
In this section, we study  the thermal partition function of the Weyl gravitino on $S^1_\beta\times S^3$ in the semi-universal limit. We will follow a similar procedure and use the same conventions for the angular velocities $(\omega_1,\omega_2)$. We will also match our result from the operator counting method. In the terminology introduced in the introduction, this theory belongs to \textit{Class 4}: it violates the naive unitarity bound and also contains negative-twist states.\\ The action of the conformal gravitino in $d=4$ dimensions is given by 
\begin{align}
S_{\frac32}
&=
\int d^4xe\,\mathcal L_{\frac32}
\end{align}
where $e\equiv\sqrt{g}$. Up to terms involving derivatives of the curvature, the Lagrangian takes the form \cite{Beccaria:2014jxa}
\begin{align}
\mathcal L_{\frac32}
&=
4ie^{-1}\epsilon^{\mu\nu\rho\sigma}
\bar\phi_\rho\gamma_5\gamma_\sigma D_\mu\phi_\nu
\nonumber\\
&\quad
+iR^{\mu\nu}
\left[
2\bar\psi^\lambda\sigma_{\lambda\nu}\phi_\mu
-2\bar\psi_\mu\sigma_{\lambda\nu}\phi^\lambda
+2\bar\psi^\lambda\gamma_\nu
\left(
D_{[\mu}\psi_{\lambda]}
-\gamma_{[\mu}\phi_{\lambda]}
\right)
\right]
-\frac{4i}{3} R\,\bar\psi^\lambda\sigma_{\lambda\nu}\phi^\nu
\end{align}
where $\phi_{\mu}$ and the covariant derivative acting on the gravitino is defined by
\begin{align}
\phi_\mu
&=
\frac13\gamma^\nu
\left(
D_\nu\psi_\mu
-D_\mu\psi_\nu
+\frac12\gamma_5\epsilon_{\nu\mu\alpha\beta}
D^\alpha\psi^\beta
\right),
\\
D_\mu\psi_\nu
&=
\left(
\partial_\mu+\frac12\sigma_{ab}\omega_\mu^{ab}(e)
\right)\psi_\nu .
\end{align}

On a Bach-flat background, and in particular on an Einstein background, one fixes the gauge symmetry by imposing 
\begin{align}
\gamma^\mu\psi_\mu=0,
\qquad
D_\mu\psi^\mu=0 .
\end{align}
Thus the physical modes are transverse and gamma-traceless. On this subspace, the quadratic Lagrangian reduces to
\begin{align}
\mathcal L_{\frac32}
&=
i\bar\psi^\lambda \mathcal O_\psi \psi_\lambda,
\\
\mathcal O_\psi
&=
\slashed D^{\,3}
+R^{\mu\nu}\gamma_\mu D_\nu
-\frac16 R\slashed D .
\end{align}

We now specialize to $S^1_\beta\times S^3$. Since the only nonzero components of the Ricci tensor are along $S^3$, we have
\begin{align}
R_{ij}
&=
\frac{R}{3}g_{ij}
=
2g_{ij},
\qquad
R=6 .
\end{align}
Therefore the kinetic operator can be written as \cite{Beccaria:2014jxa}
\begin{align}
\mathcal O_\psi
&=
\slashed D^{\,3}
+2{\slashed \nabla}_{S^3}
-\slashed D
\nonumber\\
&=
\left(
\gamma^0\partial_0+{\slashed \nabla}_{S^3}
\right)^3
-\gamma^0\partial_0
+{\slashed \nabla}_{S^3}~,
{\qquad
\slashed \nabla}_{S^3}=\gamma_iD^i .
\end{align}
Using the following relation
\begin{align}
\left\{
\slashed \nabla_{S^3},\gamma^0
\right\}
=0,
\end{align}
the determinant factorizes as
\begin{align}
\det \left(i\mathcal O^{\perp}_{\frac{3}{2}}\right)
&=
\left\{
\det\left(-\left[
\partial_0^2+\mathbf{\slashed \nabla}^{\,2}_{S^3}
\right]\right)
\det\left(-\left[
(\partial_0+1)^2+\mathbf{\slashed \nabla}^{\,2}_{S^3}
\right]\right)
\det\left(-\left[
(\partial_0-1)^2+\mathbf{\slashed \nabla}^{\,2}_{S^3}
\right]\right)
\right\}^{1/2}.
\end{align}
This is very analogous to the form of the four-derivative scalar operator: instead of two scalar towers, the conformal gravitino gives three fermionic towers.\\\\
Transverse vector-spinor harmonics $Y^{i;+}_{\ell;m_L^+,m_R^+}$ on $S^3$ belong to the representation
\begin{equation}
\left(\frac{\ell+3}{2},\frac{\ell}{2}\right)~,\quad\ell=0,1,2,\cdots
\end{equation}
with quantum numbers
\begin{equation}
m^+_L=-\frac{\ell+3}{2},\cdots,\frac{\ell+3}{2}\qquad m^+_R=-\frac{\ell}{2},\cdots,\frac{\ell}{2}~.
\end{equation}
There is also another vector-spinor harmonics  $Y^{i;-}_{\ell;m^-_L,m^-_R}$ on $S^3$ that belong to the representation
\begin{equation}
\left(\frac{\ell}{2},\frac{\ell+3}{2}\right)~,
\qquad
\ell=0,1,2,\cdots
\end{equation}
with quantum numbers
\begin{equation}
m^-_L=-\frac{\ell}{2},\cdots,\frac{\ell}{2}\qquad m^-_R=-\frac{\ell+3}{2},\cdots,\frac{\ell+3}{2}~.
\end{equation}
We impose twisted anti-periodic boundary conditions along the thermal circle $S^1_\beta$ in the following way:
\begin{equation}\label{twistedbcvector}
\psi^{i}(\tau+\beta,\Omega)
=
-e^{\,\beta\,\omega_1 J_1+\beta\,\omega_2 J_2}\,
\psi^{i}(\tau,\Omega)~.
\end{equation}
To determine the partition function, we have to sum over the spectrum of the kinetic operator. We can write down the mode expansion of the gravitino field $\psi^i$ on $S^1_\beta\times S^3$:
\begin{equation}
\psi^{i}(\tau,\Omega)
=
\sum_{n\in\mathbb Z}
\left(\sum_{\ell,m^+_L,m^+_R}e^{i\omega^+_n\tau}c^+_{n;\ell;m^+_L,m^+_R}Y^{i;+}_{\ell;m^+_L,m^+_R}(\Omega)+\sum_{\ell,m^-_L,m^-_R}e^{i\omega^-_n\tau}c^-_{n;\ell;m^-_L,m^-_R}Y^{i;-}_{\ell;m^-_L,m^-_R}(\Omega)\right).
\end{equation}
For a mode labelled by $(n,\ell)$, the twisted fermionic Matsubara
frequency is
\begin{equation}
\omega^\pm_{n}
=
\frac{2\pi}{\beta}\left(n+\frac12\right)
-i\mu^\pm(m^\pm_L,m^\pm_R),
\qquad
n\in\mathbb Z ,
\end{equation}
where
\begin{equation}
\mu^\pm(m_L,m_R)
=
(\omega_1+\omega_2)m^\pm_L
+
(\omega_1-\omega_2)m^\pm_R
\end{equation}
and $(m^\pm_L,m^\pm_R)$ take its value from which transverse vector-spinor representation $\pm$ they belong.


Using the factorized form of the determinant, we obtain\footnote{For a transverse vector-spinor on the unit $S^3$,
\begin{equation}
-\slashed{\nabla}_{S^3}^2
\quad\longrightarrow\quad
\left(\ell+\frac{5}{2}\right)^2.
\end{equation}}
\begin{align}
\det \left(i{\mathcal O}^{\perp}_{\frac{3}{2}}\right)
&=
\left(
\det\left[
\partial_0^2-\lambda_\ell^2
\right]
\det\left[
(\partial_0+1)^2-\lambda_\ell^2
\right]
\det\left[
(\partial_0-1)^2-\lambda_\ell^2
\right]
\right)^{1/2}
\nonumber\\
&=
\left(
\det\left[
\partial_0^2-\lambda_\ell^2
\right]
\det\left[
\partial_0^2-(\lambda_\ell-1)^2
\right]
\det\left[
\partial_0^2-(\lambda_\ell+1)^2
\right]
\right)^{1/2}
\end{align}
with
\begin{equation}
\lambda_\ell=\ell+\frac52~.
\end{equation}
In the final line, we have expanded the product of the determinants and reorganized them to write the final expression.

Equivalently, the three factors correspond to the three eigenvalues
\begin{align}
\lambda_\ell~,
\qquad
\lambda_\ell-1~,
\qquad
\lambda_\ell+1~.
\end{align}
Using the factorized determinant, the contribution of this mode is
\begin{equation}
\lambda^{\left(\frac{3}{2}\right)}_{n,\ell}
=
\left(\omega_{n}^2+\lambda_\ell^2\right)
\left(\omega_{n}^2+(\lambda_\ell-1)^2\right)
\left(\omega_{n}^2+(\lambda_\ell+1)^2\right)~,\qquad \ell=0,1,2,\cdots~.
\end{equation}
Therefore, using the Schwinger proper-time representation, we obtain
\begin{align}
\ln\det\left(i\mathcal{O}^{\perp}_{\frac{3}{2}}\right)
&=
-\int_0^\infty \frac{ds}{s}
\sum_{n,\ell,\pm}
\Big[
e^{-s\left(\omega_{n}^2+(\ell+\frac32)^2\right)}
+
e^{-s\left(\omega_{n}^2+(\ell+\frac52)^2\right)}
\nonumber\\
&\hspace{4.8cm}
+
e^{-s\left(\omega_{n}^2+(\ell+\frac72)^2\right)}
\Big] .
\end{align}
Here also we have multiplied by an extra $2$ factor because we have considered complex vector fermion.

Repeating the same Poisson resummation as before, and dropping the zero-winding
Casimir contribution, we get the thermal partition function
\begin{align}
\ln\det\left(i\mathcal{O}^{\perp}_{\frac{3}{2}}\right)
&=
2\sum_{k=1}^{\infty}\sum_{\ell=0}^{\infty}
\frac{(-1)^{k+1}}{k}
\left(
e^{-k\beta(\ell+\frac32)}
+
e^{-k\beta(\ell+\frac52)}
+
e^{-k\beta(\ell+\frac72)}
\right)\\&~~~~~\times \Bigg[\sum_{m^+_L,m^+_R}e^{k\beta\mu^+}+\sum_{m^-_L,m^-_R}e^{k\beta\mu^-}\Bigg]~.
\end{align}
As before, the sum over $m^\pm_L,m^\pm_R$ can be written as the corresponding character sum,
\begin{equation}
\begin{split}
&\chi^{(\frac32)}_\ell(k;\omega_1,\omega_2)=\Bigg[\sum_{m^+_L,m^+_R}e^{k\beta\mu^+}+\sum_{m^-_L,m^-_R}e^{k\beta\mu^-}\Bigg]\\&=\frac{
\sinh\!\left(\frac{k\beta(\ell+4)(\omega_1+\omega_2)}{2}\right)
}{
\sinh\!\left(\frac{k\beta(\omega_1+\omega_2)}{2}\right)
}
\frac{
\sinh\!\left(\frac{k\beta(\ell+1)(\omega_1-\omega_2)}{2}\right)
}{
\sinh\!\left(\frac{k\beta(\omega_1-\omega_2)}{2}\right)
}+\frac{
\sinh\!\left(\frac{k\beta(\ell+1)(\omega_1+\omega_2)}{2}\right)
}{
\sinh\!\left(\frac{k\beta(\omega_1+\omega_2)}{2}\right)
}
\frac{
\sinh\!\left(\frac{k\beta(\ell+4)(\omega_1-\omega_2)}{2}\right)
}{
\sinh\!\left(\frac{k\beta(\omega_1-\omega_2)}{2}\right)
}~.    
\end{split}
\end{equation}
Therefore, we can rewrite the partition function as
\begin{align}
\ln\det\left(i\mathcal{O}^{\perp}_{\frac{3}{2}}\right)
&=
2\sum_{k=1}^{\infty}
\frac{(-1)^{k+1}}{k}
\sum_{\ell=0}^{\infty}
\left(
e^{-k\beta(\ell+\frac32)}
+
e^{-k\beta(\ell+\frac52)}
+
e^{-k\beta(\ell+\frac72)}
\right)
\chi^{(\frac32)}_\ell(k;\omega_1,\omega_2).
\end{align}
Writing
\begin{equation}
q=e^{-k\beta},
\qquad
x_L=e^{\frac{k\beta}{2}(\omega_1+\omega_2)},
\qquad
x_R=e^{\frac{k\beta}{2}(\omega_1-\omega_2)},
\end{equation}
the remaining $\ell$-sum can be performed exactly. One finds
\begin{align}
&\sum_{\ell=0}^{\infty}
\left(
q^{\ell+\frac32}
+
q^{\ell+\frac52}
+
q^{\ell+\frac72}
\right)
\chi^{(\frac32)}_\ell(k;\omega_1,\omega_2)
\nonumber\\
&\qquad =
\frac{2
q^{\frac32}(1+q+q^2)\,
\mathcal N'_{\frac32}(q;x_L,x_R)
}
{
(1-qe^{k\beta\omega_1})(1-qe^{-k\beta\omega_1})
(1-qe^{k\beta\omega_2})(1-qe^{-k\beta\omega_2})
}~,
\end{align}
where
\begin{align}
\mathcal N'_{\frac32}(q;x_L,x_R)
&=
\chi_{\frac32}(x_L)+\chi_{\frac32}(x_R)
\nonumber\\
&\quad
-q\left[
\chi_1(x_L)\chi_{\frac12}(x_R)
+
\chi_{\frac12}(x_L)\chi_1(x_R)
\right]
+q^2\left[
\chi_{\frac12}(x_L)+\chi_{\frac12}(x_R)
\right]~.
\end{align}
We also have to add $\det^{\prime}\left(i\mathcal{O}^{\perp}_{\frac{1}{2}}\right)$ that comes from the fermion where ``prime'' represents the $\ell=0$ mode has been subtracted. Performing the $\ell$ sum, we can write 
\begin{equation}
\begin{split}
\ln{\det}^{\prime}\left(i\mathcal{O}_{\frac{1}{2}}\right)&=\sum_{k=1}^{\infty}\frac{(-1)^{k+1}}{k}\frac{2
q^{\frac32}\,
\mathcal N''_{\frac12}(q;x_L,x_R)
}
{
(1-qe^{k\beta\omega_1})(1-qe^{-k\beta\omega_1})
(1-qe^{k\beta\omega_2})(1-qe^{-k\beta\omega_2})
}    
\end{split}    
\end{equation}
where ${\mathcal N}''_{\frac12}(q;x_L,x_R)$ is given by
\begin{equation}
\begin{split}
&{\mathcal N}''_{\frac12}(q;x_L,x_R)=-q^2\left[\chi_{\frac32}(x_L)+\chi_{\frac32}(x_R)\right]
\\
&\quad
+q(1-q+q^2)\left[
\chi_1(x_L)\chi_{\frac12}(x_R)
+
\chi_{\frac12}(x_L)\chi_1(x_R)
\right]
-q^2(1-q+q^2)\left[
\chi_{\frac12}(x_L)+\chi_{\frac12}(x_R)
\right]~.    
\end{split}    
\end{equation}
The full partition function $Z_{\frac{3}{2}}$ for Weyl gravitino \cite{Beccaria:2014xda} is given by
\begin{equation}
Z_{\frac{3}{2}}=\det \left(i{\mathcal O}^{\perp}_{\frac{3}{2}}\right){\det}^{\prime}\left(i\mathcal{O}_{\frac{1}{2}}\right)    
\end{equation}
which can be rewritten as a sum over single-particle partition function $z_{\frac{3}{2}}(k)$
\begin{align}
\ln Z_{\frac32}
&=
\sum_{k=1}^{\infty}
\frac{(-1)^{k+1}}{k}\,
z_{\frac32}(k),
\end{align}
with the refined single-particle partition function
\begin{equation}
\begin{split}
&z_{\frac32}(k)
=
\frac{
2q^{\frac{3}{2}}\mathcal{N}_{\frac{3}{2}}(q,x_L,x_R)
}
{
(1-qe^{k\beta\omega_1})(1-qe^{-k\beta\omega_1})
(1-qe^{k\beta\omega_2})(1-qe^{-k\beta\omega_2})
}\\ &\mathcal{N}_{\frac{3}{2}}(q,x_L,x_R)=\Bigg[(1+q)\left(\chi_{\frac{3}{2}}(x_L)+\chi_{\frac{3}{2}}(x_R)\right)-2q^2\left(\chi_{\frac{1}{2}}(x_L)\chi_1(x_R)+\chi_{1}(x_L)\chi_{\frac{1}{2}}(x_R)\right)\\&~~~~~~~~~~~~~~~~~~~~~~~~+2q^3\left(\chi_{\frac{1}{2}}(x_L)+\chi_{\frac{1}{2}}(x_R)\right)
\Bigg]~.
\end{split}
\end{equation}
The partition function of the free Weyl gravitino field can also be obtained using the operator counting method. The gauge invariant primary operator have scaling dimension $\Delta=\frac{3}{2}$ which is gravitino curvature $\psi_{\mu\nu}\equiv D_\mu\psi_\nu-D_\nu\psi_\mu$. The the covariant Weyl tensor field strength $\mathrm{\Phi}_{\mu\nu}$ for the conformal gravitino is given by 
\begin{equation}
\mathrm{\Phi}_{\mu\nu}\equiv\frac{1}{3}\left(\psi_{\mu\nu}-\gamma_5\tilde{\psi}_{\mu\nu}+2\gamma_{[\nu}^{~\lambda}\psi_{\lambda\mu]}\right)    
\end{equation}
and they obey
\begin{equation}
\mathrm{\Phi}_{\mu\nu}=\gamma_5\tilde{\mathrm{\Phi}}_{\mu\nu}~,\qquad\gamma^\mu \mathrm{\Phi}_{\mu\nu}=0    
\end{equation}
where ``tilde'' implies the Hodge dual. This field strength decomposes into the self-dual and the anti self-dual pieces, which transform as $\left(\frac{3}{2},0\right)\oplus\left(0,\frac{3}{2}\right)$. Therefore, the spin character of this primary is
$$q^{\frac{3}{2}}\left[\chi_{\frac{3}{2}}(x_L)+\chi_{\frac{3}{2}}(x_R)\right]$$
where $x_L=e^{k\beta\frac{(\omega_1+\omega_2)}{2}}$ and $x_R=e^{k\beta\frac{(\omega_1-\omega_2)}{2}}$. The similar thing happens for $\slashed{D}\mathrm{\Phi}_{\mu\nu}$ which contribute
$$q^{\frac{5}{2}}\left[\chi_{\frac{3}{2}}(x_L)+\chi_{\frac{3}{2}}(x_R)\right]~.$$
The unrestricted descendant module is obtained by acting with arbitrary derivatives on the primary. Since, the derivative transforms as $\left(\frac{1}{2},\frac{1}{2}\right)$, and each derivative contributes to either of the four weights,i.e, $qx_1$, $qx^{-1}_1$, $q x_{2}$ or $qx^{-1}_2$, the unrestricted descendant module gives 
$$\frac{1}{\left(1-q\, x_1\right)\left(1-q\, x^{-1}_1\right)\left(1-q\, x_2\right)\left(1-q\, x^{-1}_2\right)}~.$$
The equations motion and the Bianchi identity involving $D^\mu\slashed{D}\mathrm{\Phi}_{\mu\nu}$ and $D^\mu\slashed{D}\tilde{\mathrm{\Phi}}_{\mu\nu}$ transform like $\left(\frac{1}{2},1\right)\oplus\left(1,\frac{1}{2}\right)$ which subtracts $$2q^{\frac{7}{2}}\left[\chi_{\frac{1}{2}}(x_L)\chi_1(x_R)+\chi_{1}(x_L)\chi_{\frac{1}{2}}(x_R)\right]$$ 
at two derivative level. We also have to add
$$2q^{\frac{9}{2}}\left[\chi_{\frac{1}{2}}(x_L)+\chi_{\frac{1}{2}}(x_R)\right]$$ 
to compensate the overcounting due to fact that
\begin{equation}
D^\mu D^\nu\slashed{D}\mathrm{\Phi}_{\mu\nu}=0~,\qquad D^\mu D^\nu\slashed{D}\tilde{\mathrm{\Phi}}_{\mu\nu}=0
\end{equation}
at the three derivative level. In this case, the left hand side of the above equations transform like $\left(\frac{1}{2},0\right)\oplus\left(0,\frac{1}{2}\right)$. So the one particle partition function for the gravitino field from the operator counting method is given by
\begin{equation}
z_{\frac{3}{2}}=\frac{
2q^{\frac32}\,
\mathcal N_{\frac32}(q;x_L,x_R)
}
{
(1-qe^{k\beta\omega_1})(1-qe^{-k\beta\omega_1})
(1-qe^{k\beta\omega_2})(1-qe^{-k\beta\omega_2})
}~.    
\end{equation}
Here also, we have multiplied by an extra $2$ because we have considered complex vector fermion.


We will study the partition function in the semi-universal limit \eqref{limit}. In that limit, our thermal partition function behaves like
\begin{equation}
\ln Z_{\frac{3}{2}}=\frac{\beta^2h_{\frac{3}{2}}(\beta)}{\nu_1\nu_2}+O\left(\frac{1}{\nu_i}\right)    
\end{equation}
where the function $h_{\frac{3}{2}}(\beta)$ is given by
\begin{equation}
h_{\frac{3}{2}}(\beta)=\frac{4}{\beta^2}\sum_{k=1}^{\infty}\frac{(-1)^{k+1}\cosh^2\left(k\beta\right)}{k^3\sinh\left(\frac{k\beta}{2}\right)}~.    
\end{equation}

Note that the residue function grows as $(-1)^{k+1}\frac{e^{\frac{3k\beta}{2}}}{k^3}$ is in the large $k$ regime of the sum. This is due to the presence of a negative twist state with twist $\tau=-\frac{3}{2}$. However, the high-temperature expansion of the partition function fails to capture the presence of the negative twist sector, but the semi-universal limit does. We discuss more about this in section \ref{ansc}.

\section{Semi-universality of the Weyl graviton on $S^1_\beta \times S^3$}
\label{grav}

We will study the  thermal partition function of the conformal spin-$2$ field (Weyl graviton) on
$S^1_\beta\times S^3$ in the semi-universal limit.
In the terminology introduced in the introduction, this theory belongs to \textit{Class 4}: it violates the naive unitarity bound and also contains negative-twist states.

The one-loop partition function is determined by the ratio of determinants \cite{Beccaria:2014jxa}
\begin{equation}
Z_2=
\frac{1}
{\left[\det \mathcal O^{\perp}_{2}\,
\det{}' \mathcal O^{\perp}_{1}\right]^{1/2}},
\end{equation}
where $\mathcal O^{\perp}_{2}$ acts on transverse traceless rank-$2$ tensors, while $\mathcal O^{\perp}_{1}$ is the transverse vector ghost operator. The ``prime'' in the determinant of $\det{}' \mathcal O^{\perp}_{1}$ implies that the zero mode in the vector Laplacian on $S^3$ needs to be dropped. Since it satisfies $D_iV_j+D_jV_i=0$, it does not appear in the decomposition of $h_{ij}$.
\begin{equation}
h^{ij}(\tau,\Omega)
=
\sum_{n\in\mathbb Z}
\left(\sum_{\ell,m^+_L,m^+_R}e^{i\omega^+_n\tau}c^+_{n;\ell;m^+_L,m^+_R}Y^{ij;+}_{\ell;m^+_L,m^+_R}(\Omega)+\sum_{\ell,m^-_L,m^-_R}e^{i\omega^-_n\tau}c^-_{n;\ell;m^-_L,m^-_R}Y^{ij;-}_{\ell;m^-_L,m^-_R}(\Omega)\right).
\end{equation}

On $S^1_\beta\times S^3$, we can express $\mathcal{O}^{\perp}_2$ operator as
\begin{equation}
\mathcal{O}^{\perp}_2=\Big[(\partial_0-1)^2+\nabla^2_{S^3}-3\Big]\Big[(\partial_0+1)^2+\nabla^2_{S^3}-3\Big]    
\end{equation}
\\ Transverse traceless rank-two tensor harmonics $Y^{ij;+}_{\ell;m^+_L,m^+_R}$ on $S^3$ belong to the representation
\begin{equation}
\left(\frac{\ell+4}{2},\frac{\ell}{2}\right)~,
\qquad
\ell=0,1,2,\cdots
\end{equation}
with quantum numbers
\begin{equation}
m^+_L=-\frac{\ell+4}{2},\cdots,\frac{\ell+4}{2}\qquad m^+_R=-\frac{\ell}{2},\cdots,\frac{\ell}{2}~.
\end{equation}
Also there is another transverse traceless rank-two tensor spherical harmonics $Y^{ij;-}_{\ell;m^-_L,m^-_R}$ on $S^3$ that belong to the representation
\begin{equation}
\left(\frac{\ell}{2},\frac{\ell+4}{2}\right)~,
\qquad
\ell=0,1,2,\cdots
\end{equation}
with quantum numbers
\begin{equation}
m^-_L=-\frac{\ell}{2},\cdots,\frac{\ell}{2}\qquad m^-_R=-\frac{\ell+4}{2},\cdots,\frac{\ell+4}{2}~.
\end{equation}
We impose twisted periodic boundary conditions along the thermal circle $S^1_\beta$ in the following way:
\begin{equation}\label{twistedbcvector}
h^{ij}(\tau+\beta,\Omega)
=
e^{\,\beta\,\omega_1 J_1+\beta\,\omega_2 J_2}\,
h^{ij}(\tau,\Omega)~.
\end{equation}
To compute the partition function, we sum over the eigenspectrum of the kinetic operator. We therefore begin by writing the mode expansion of the transverse traceless graviton field on $S^1_{\beta}\times S^3$:
\begin{equation}
h^{ij}(\tau,\Omega)
=
\sum_{n\in\mathbb Z}
\left(\sum_{\ell,m^+_L,m^+_R}e^{i\omega^+_n\tau}c^+_{n;\ell;m^+_L,m^+_R}Y^{ij;+}_{\ell;m^+_L,m^+_R}(\Omega)+\sum_{\ell,m^-_L,m^-_R}e^{i\omega^-_n\tau}c^-_{n;\ell;m^-_L,m^-_R}Y^{ij;-}_{\ell;m^-_L,m^-_R}(\Omega)\right).
\end{equation}
The periodic boundary conditions implies
\begin{equation}
\omega^{\pm}_n=\frac{2n\pi}{\beta}-i\left[(\omega_1+\omega_2)m^\pm_L+(\omega_1-\omega_2)m^\pm_R\right]~,\qquad n\in\mathbb{Z}    
\end{equation}
and $(m^\pm_L,m^\pm_R)$ take its value from which transverse traceless rank-two tensor representation $\pm$ they belong.\\
The complete set of eigenvalues of $\mathcal O^{\perp}_2$ are given by: \footnote{For a rank-two tensor on the unit $S^3$,
\begin{equation}
-\nabla_{S^3}^2+3
\quad\longrightarrow\quad
\left(\ell+3\right)^2.
\end{equation}}
\begin{equation}
\lambda^{(2)}_{n,\ell}
=
\bigl(\omega_n^2+(\ell+2)^2\bigr)
\bigl(\omega_n^2+(\ell+4)^2\bigr)~,
\qquad \ell=0,1,2,\cdots~.
\end{equation}




Repeating the same analysis involving proper-time and Poisson resummation as done in the scalar examples, one obtains
\begin{equation}
\begin{split}
-\frac{1}{2}\ln\det\mathcal{O}^\perp_2=\sum_{k=1}^{\infty}\sum_{\ell=0}^{\infty}\frac{2\cosh(k\beta)e^{-k\beta(\ell+3)}}{k} \Bigg[\sum_{m^+_L,m^+_R}e^{k\beta\mu^+}+\sum_{m^-_L,m^-_R}e^{k\beta\mu^-}\Bigg]~.       
\end{split}    
\end{equation}
Here also, we can express $(m^{\pm}_L,m^{\pm}_R)$ sums as the product of two $SU(2)$ characters. So we have
\begin{equation}
\sum_{m^+_L,m^+_R}e^{k\beta\mu^+(m_L^+,m_R^+)}=\frac{
\sinh\!\left(\frac{k\beta(\ell+5)(\omega_1+\omega_2)}{2}\right)
}{
\sinh\!\left(\frac{k\beta(\omega_1+\omega_2)}{2}\right)
}
\frac{
\sinh\!\left(\frac{k\beta(\ell+1)(\omega_1-\omega_2)}{2}\right)
}{
\sinh\!\left(\frac{k\beta(\omega_1-\omega_2)}{2}\right)
}
\end{equation}
and similarly we have
\begin{equation}
\sum_{m^-_L,m^-_R}e^{k\beta\mu^-(m^-_L,m^-_R)}=\frac{
\sinh\!\left(\frac{k\beta(\ell+1)(\omega_1+\omega_2)}{2}\right)
}{
\sinh\!\left(\frac{k\beta(\omega_1+\omega_2)}{2}\right)
}
\frac{
\sinh\!\left(\frac{k\beta(\ell+5)(\omega_1-\omega_2)}{2}\right)
}{
\sinh\!\left(\frac{k\beta(\omega_1-\omega_2)}{2}\right)
}~.
\end{equation}
Performing the $\ell$ sum, we can write
\begin{equation}
\begin{split}
&-\frac{1}{2}\ln\det\mathcal{O}^\perp_2\\&=\sum_{k=1}^{\infty}\frac{1}{k}\frac{2e^{-2k\beta}\left(1+e^{-2k\beta}\right)}{\left(1-e^{-k\beta(1-\omega_1)}\right)\left(1-e^{-k\beta(1+\omega_1)}\right)
 \left(1-e^{-k\beta(1-\omega_2)}\right)\left(1-e^{-k\beta(1+\omega_2)}\right)}\\&~~~~\times\Big[1+2\cosh(k\beta\omega_1)\cosh(k\beta\omega_2)+2\cosh(2k\beta\omega_1)\cosh(2k\beta\omega_2)\\&~~~~~~~-2e^{-k\beta}\big(\cosh(2k\beta\omega_1)\cosh(k\beta\omega_2)+\cosh(k\beta\omega_1)\cosh(2k\beta\omega_2)+\cosh(k\beta\omega_1)+\cosh(k\beta\omega_2)\big)\\&~~~~~~~~+e^{-2k\beta}\big(1+2\cosh(k\beta\omega_1)\cosh(k\beta\omega_2)\big)\Big]~.    
\end{split}    
\end{equation} 
We also need to compute $\ln{\det}'\mathcal{O}^\perp_1 $. To do that we will use \eqref{partitionfunctionvector} but perform the sum from $\ell=1$ to $\infty$ instead of $0$ to $\infty$. The result is given by
\begin{equation}
\begin{split}
&-\frac{1}{2}\ln{\det}'\mathcal{O}^\perp_1\\&=\sum_{k=1}^{\infty}\frac{1}{k}\frac{2e^{-3k\beta}}{\left(1-e^{-k\beta(1-\omega_1)}\right)\left(1-e^{-k\beta(1+\omega_1)}\right)
 \left(1-e^{-k\beta(1-\omega_2)}\right)\left(1-e^{-k\beta(1+\omega_2)}\right)}\\&~~~~~~~~~\times\Big[4\cosh(k\beta\omega_1)\cosh(k\beta\omega_2)\big(\cosh(k\beta\omega_1)+\cosh(k\beta\omega_2)\big)\\&~~~~~~~~~~+e^{-k\beta}\big(1-2\big(1+2\cosh(k\beta\omega_1)\cosh(k\beta\omega_2)\big)^2\big)\\&~~~~~~~~~~+2e^{-2k\beta}\big(\cosh(k\beta\omega_1)+\cosh(k\beta\omega_2)\big)\big(1+2\cosh(k\beta\omega_1)\cosh(k\beta\omega_2)\big)\\&~~~~~~~~~~-e^{-3k\beta}\big(1+2\cosh(k\beta\omega_1)\cosh(k\beta\omega_2)\big)\Big]~. 
\end{split}    
\end{equation}
So the full partition function is given by
\begin{equation}
\ln Z_2=-\frac{1}{2}\ln\det\mathcal{O}^\perp_2-\frac{1}{2}\ln{\det}'\mathcal{O}^\perp_1
\end{equation}
which we can write as
\begin{equation}
\ln Z_2
=
\sum_{k=1}^{\infty}\frac1k\,z_2(k),
\end{equation}
where the refined one-particle partition function is
\begin{equation}\label{modesumgrav}
\begin{split}
z_2(k)
&=
\frac{2e^{-2k\beta}
}{
\left(1-e^{-k\beta(1-\omega_1)}\right)
\left(1-e^{-k\beta(1+\omega_1)}\right)
\left(1-e^{-k\beta(1-\omega_2)}\right)
\left(1-e^{-k\beta(1+\omega_2)}\right)
}\\&\times\Big[
1+2\cosh(k\beta\omega_1)\cosh(k\beta\omega_2)
+2\cosh(2k\beta\omega_1)\cosh(2k\beta\omega_2)\\
&~~~~~~~-e^{-2k\beta}
\big(1+
4\cosh(k\beta\omega_1)\cosh(k\beta\omega_2)
+2\cosh(2k\beta\omega_1)+2\cosh(2k\beta\omega_2)
\big)\\
&~~~~~~~~~~~~~+2e^{-3k\beta}
\big(\cosh(k\beta\omega_1)
+\cosh(k\beta\omega_2)\big)\Big]~.
\end{split}
\end{equation}

The partition function of the Weyl graviton can be obtained easily using operator counting method. For Weyl graviton, the natural gauge invariant operator is the linearized Weyl tensor. In spinor notation, it comes with two chiral pairs: the self dual $C_{\alpha\beta\gamma\delta}$ and the anti-self-dual $C_{\dot\alpha\dot\beta\dot\gamma\dot\delta}$. Taken together, the Weyl graviton transforms as $(2,0)\oplus (0,2)$, with both components having the conformal dimension $\Delta=2$.

We now have to remove the null states corresponding to the Bach equation of motions
\begin{equation}\label{Bach}
\partial^{\alpha}{}_{\dot\alpha}
\partial^{\beta}{}_{\dot\beta}
C_{\alpha\beta\gamma\delta}=0 ,\qquad \partial_{\alpha}{}^{\dot\alpha}
\partial_{\beta}{}^{\dot\beta}
C_{\dot\alpha\dot\beta\dot\gamma\dot\delta}=0 .
\end{equation}
Both of these equations can be thought of as transformations under $\left(1,1\right)$.
These Bach equations are also divergenceless,
\begin{equation}
\partial^{\gamma\dot\alpha}
\partial^{\alpha}{}_{\dot\alpha}
\partial^{\beta}{}_{\dot\beta}
C_{\alpha\beta\gamma\delta}=0, \qquad\partial^{\alpha\dot\gamma} \partial_{\alpha}{}^{\dot\alpha}
\partial_{\beta}{}^{\dot\beta}
C_{\dot\alpha\dot\beta\dot\gamma\dot\delta}=0,
\end{equation}
and these equations transform under $\left(\frac{1}{2},\frac{1}{2}\right)$

Therefore, removing the null states, the single particle partition function can be written as
\begin{align}\label{opcountgrav}
    z_2(k)&=\frac{q^2 \left(\chi_2(x_L)+\chi_2(x_R)\right)-2 q^4\left(\chi_1(x_L)\times \chi_1(x_R)\right)+2 q^5 \left(\chi_\frac{1}{2}(x_L)\times \chi_\frac{1}{2}(x_R)\right)}{(1-q x_1)(1-qx^{-1}_1)(1-q x_2)(1-q x^{-1}_2)}.
\end{align}
Here $\chi_j(x)$ is the spin-$j$ $SU(2)$ character. After explicit substitution of $q=e^{-k\beta}$ and $x_i=e^{k\beta\omega_i}$, we find agreement with \eqref{modesumgrav} and \eqref{opcountgrav}.



Let us now study the partition function in the semi-universal limit \eqref{limit}. In this limit, it becomes
\begin{equation}
\ln Z_2=\frac{\beta^2h_2(\beta)}{\nu_1\nu_2}+O\left(\frac{1}{\nu_i}\right)    
\end{equation}
where the function $h_2(\beta)$ is given by
\begin{equation}
h_2(\beta)=\frac{1}{\beta^2}\sum_{k=1}^{\infty}\frac{\coth\left(k\beta\right)\left(3+4\sinh^2\left(k\beta\right)\right)}{k^3}~.    
\end{equation}

We note that the residue function is not well behaved as it grows like $\frac{e^{2k\beta}}{k^3}$ for the large $k$-terms in the sum. This phenomenon can again be attributed to the presence of a negative twist state, with $\tau=-2$, in the single particle partition function. Therefore, the theory of the Weyl graviton, which is expected to violate the ANEC-type bound, also exhibits pathological behavior in the semi-universal limit. However, the traditional high-temperature expansion is completely insensitive to this behavior. We discuss more about this in section \ref{ansc}.
\section{Semi-universality of conformal higher spins on $S^1_{\beta}\times S^3$}
We will study the  thermal partition function of the conformal spin-$s$ field on
$S^1_\beta\times S^3$ in the semi-universal limit.
In the terminology introduced in the introduction, this theory belongs to \textit{Class 4}: it violates the naive unitarity bound and also contains negative-twist states.
\subsection{Boundary partition function}\label{chsbdy} 
The partition functions of conformal higher-spin theories  are objects of considerable interest, especially for testing holographic dualities in various dimensions \cite{Giombi:2013yva,Tseytlin:2013jya}.
In this section, we study the partition function of conformal higher spin fields on $S^1_{\beta}\times S^3$, with two angular momenta $\omega_1$ and $\omega_2$ corresponding to the two planes of rotation. We will investigate the partition function in the limit $\omega_i \rightarrow 1$.

We consider a conformal spin-$s$ field on $S^1_\beta \times S^3$ whose action contains
$2s$ derivatives. Restricting to transverse-traceless tensor modes
$\phi_{i_1 i_2 \cdots i_s}$ on $S^3$, the kinetic operator takes the following form for even and odd spins, respectively \cite{Beccaria:2014jxa} 
\begin{equation}
\mathcal{O}^{\perp}_s=\begin{cases}

&
\prod_{p=1}^{s}
\big[
(\partial_0+2p-s-1)^2+\nabla^2_{S^3}-(s+1)
\big]~,\qquad s={\rm even}~,
\\[1.6em]
&
-
\prod_{p=-\frac{s-1}{2}}^{\frac{s-1}{2}}
\left[
(\partial_0+2p)^2+\nabla^2_{S^3}-(s+1)
\right]~,\qquad s={\rm odd}~.    
\end{cases}
\end{equation}
Spin $s$-tensor harmonics belongs to the representation 
\begin{equation}
\left(\frac{\ell+2s}{2},\frac{\ell}{2}\right)~,\quad\textmd{where}~~\ell=0,1,2,\cdots   
\end{equation} with the quantum numbers 
\begin{equation}
m^+_L=-\frac{\ell+2s}{2},\cdots,\frac{\ell+2s}{2}\qquad m^+_R=-\frac{\ell}{2},\cdots,\frac{\ell}{2}~.
\end{equation}
Or in the representation
\begin{equation}
\left(\frac{\ell}{2},\frac{\ell+2s}{2}\right)~,\quad\textmd{where}~~\ell=0,1,2,\cdots   
\end{equation} with the quantum numbers 
\begin{equation}
m^-_L=-\frac{\ell}{2},\cdots,\frac{\ell}{2}\qquad m^-_R=-\frac{\ell+2s}{2},\cdots,\frac{\ell+2s}{2}~.
\end{equation}
The periodic boundary conditions imply
\begin{equation}
\omega^{\pm}_n=\frac{2n\pi}{\beta}-i\left[(\omega_1+\omega_2)m^\pm_L+(\omega_1-\omega_2)m^\pm_R\right]~,\qquad n\in\mathbb{Z}    
\end{equation}
and $(m^\pm_L,m^\pm_R)$ takes its value from the tensor representation $\pm$ to which they belong.\\
The complete set of eigenvalues of $\mathcal O^{\perp}_s$ are given by: \footnote{For a general tensor on the unit $S^3$ the eigenvalues of the kinetic operator are given by,
\begin{equation}
\left(-\nabla_{S^3}^2+s+1\right)\psi_s(\Omega)
=
\left(\ell+s+1\right)^2\psi_s(\Omega).
\end{equation}}
\begin{equation}
\begin{split}
\lambda^{(s)}_{n,\ell}=\prod_{p=1}^{s}
\left[
\omega^2_n+(\ell+2p)^2
\right]~.    
\end{split}    
\end{equation}
The proposed partition function $Z^{\rm CHS}_s$ is given by
\begin{equation}
Z^{\rm CHS}_s=\frac{1}{\left[\prod_{p=1}^{s}\det^{\prime}\mathcal{O}^{\perp}_p\right]^{1/2}}    
\end{equation}
where the ``prime'' in the determinant implies that the product of the eigenvalues should start from $\ell=s-p$. By performing the proper-time parameterization and Poisson resummation, and also dropping out the Casimir energy, we can express the logarithm of the partition function\footnote{Here also $\mu_p^{\pm}=(\omega_1+\omega_2)m^\pm_L+(\omega_1-\omega_2)m^\pm_R$.} as
\begin{equation}
 \begin{split}
 \ln Z^{\rm CHS}_{p,s-p}=-\frac{1}{2}\ln{\det}^{\prime}\mathcal{O}^{\perp}_p=\sum_{k=1}^{\infty}\frac{1}{k}\sum_{r=1}^{p}\sum_{\ell=s-p}^{\infty}e^{-k\beta(\ell+2r)}\Bigg[\sum_{m^+_L,m^+_R}e^{k\beta\mu^+_p}+\sum_{m^-_L,m^-_R}e^{k\beta\mu^-_p}\Bigg]~.    
 \end{split}   
\end{equation}
Here also, we can express $(m^{\pm}_L,m^{\pm}_R)$ sums as the product of two $SU(2)$ characters. So we have
\begin{equation}
\sum_{m^+_L,m^+_R}e^{k\beta\mu^+_p(m_L^+,m_R^+)}=\frac{
\sinh\!\left(\frac{k\beta(\ell+2p+1)(\omega_1+\omega_2)}{2}\right)
}{
\sinh\!\left(\frac{k\beta(\omega_1+\omega_2)}{2}\right)
}
\frac{
\sinh\!\left(\frac{k\beta(\ell+1)(\omega_1-\omega_2)}{2}\right)
}{
\sinh\!\left(\frac{k\beta(\omega_1-\omega_2)}{2}\right)
}
\end{equation}
and similarly we have
\begin{equation}
\sum_{m^-_L,m^-_R}e^{k\beta\mu^-_p(m^-_L,m^-_R)}=\frac{
\sinh\!\left(\frac{k\beta(\ell+1)(\omega_1+\omega_2)}{2}\right)
}{
\sinh\!\left(\frac{k\beta(\omega_1+\omega_2)}{2}\right)
}
\frac{
\sinh\!\left(\frac{k\beta(\ell+2p+1)(\omega_1-\omega_2)}{2}\right)
}{
\sinh\!\left(\frac{k\beta(\omega_1-\omega_2)}{2}\right)
}~.
\end{equation}
Performing $\ell$ sum and $p$ sum, we can express have log of the partition function as
\begin{equation}
\begin{split}
\ln Z^{\rm CHS}_{p,s-p}=\sum_{k=1}^{\infty}\frac{1}{k}z_{p,s-p}(k)    
\end{split}    
\end{equation}
where the single particle partition function $z_{p,s-p}(k)$ is given by
\begin{equation}
\begin{split}
z_{p,s-p}(k)=\frac{q^{s+2-p}\left(1-q^{2p}\right)\mathcal{N}_{s,p}(q,x_L,x_R)}{\left(1-q^2\right)(1-qx_1)(1-qx^{-1}_1)(1-qx_2)(1-qx^{-1}_2)}    
\end{split}    
\end{equation}
where $q=e^{-k\beta},~x_1=e^{k\beta\omega_1},~x_2=e^{k\beta\omega_2}$ and $x_L=(x_1x_2)^{\frac{1}{2}},~x_R=\left(\frac{x_1}{x_2}\right)^{\frac{1}{2}}$. The expression of $\mathcal{N}_{s,p}(q,x_L,x_R)$ is given by
\begin{equation}
\begin{split}
&\mathcal{N}_{s,p}(q,x_L,x_R)\\&=
\chi_{\frac{s+p}{2}}(x_L)\chi_{\frac{s-p}{2}}(x_R)
+
\chi_{\frac{s-p}{2}}(x_L)\chi_{\frac{s+p}{2}}(x_R)
\\
&
-q
\Big[
\chi_{\frac{s+p+1}{2}}(x_L)\chi_{\frac{s-p-1}{2}}(x_R)
+\chi_{\frac{s+p-1}{2}}(x_L)\chi_{\frac{s-p+1}{2}}(x_R)+
\chi_{\frac{s-p+1}{2}}(x_L)\chi_{\frac{s+p-1}{2}}(x_R)
\\&~~~~~~~+\chi_{\frac{s-p-1}{2}}(x_L)\chi_{\frac{s+p+1}{2}}(x_R)+
\chi_{\frac{s+p-1}{2}}(x_L)\chi_{\frac{s-p-1}{2}}(x_R)+\chi_{\frac{s-p-1}{2}}(x_L)\chi_{\frac{s+p-1}{2}}(x_R)
\Big]
\\
&+q^2
\Big[
\chi_{\frac{s+p-2}{2}}(x_L)\chi_{\frac{s-p}{2}}(x_R)
+\chi_{\frac{s+p}{2}}(x_L)\chi_{\frac{s-p-2}{2}}(x_R)+
\chi_{\frac{s-p-2}{2}}(x_L)\chi_{\frac{s+p}{2}}(x_R)
\\&~~~~~~~+\chi_{\frac{s-p}{2}}(x_L)\chi_{\frac{s+p-2}{2}}(x_R)+
\chi_{\frac{s+p}{2}}(x_L)\chi_{\frac{s-p}{2}}(x_R)
+
\chi_{\frac{s-p}{2}}(x_L)\chi_{\frac{s+p}{2}}(x_R)
\Big]\\&-q^3\Big[\chi_{\frac{s+p-1}{2}}(x_L)\chi_{\frac{s-p-1}{2}}(x_R)
+
\chi_{\frac{s-p-1}{2}}(x_L)\chi_{\frac{s+p-1}{2}}(x_R)\Big]~.
\end{split}    
\end{equation}
For boundary cases such as $s=p$, terms like $\chi_{-\frac{1}{2}}(x)$ or $\chi_{-1}(x)$ should be understood by analytic continuation of $\chi_{j}(x)$; in particular $\chi_{-\frac{1}{2}}(x)=0,~\chi_{-1}(x)=-1$ etc.

So the full log partition function can be expressed as
\begin{equation}
\ln Z^{\rm CHS}_s=\sum_{p=1}^s\ln Z_{p,s-p}=\sum_{k=1}^{\infty}\frac{1}{k}z_s(k)    
\end{equation}
where the single particle partition function $z_s(k)$ is given by
\begin{equation}\label{partitions}
\begin{split}
z_s(k)&=\sum_{p=1}^sz_{p,s-p}(k)\\&=\frac{q^2\left[\chi_s(x_L)+\chi_s(x_R)-2q^s\chi_{\frac{s}{2}}(x_L)\chi_{\frac{s}{2}}(x_R)+2q^{s+1}\chi_{\frac{s-1}{2}}(x_L)\chi_{\frac{s-1}{2}}(x_R)\right]}{(1-qx_1)(1-qx^{-1}_1)(1-qx_2)(1-qx^{-1}_2)}~.    
\end{split}    
\end{equation}

The partition function of the conformal higher spin theory can also be obtained from operator counting method. In this theory, the natural gauge invariant operator is generalized Weyl tensor field $C_{\mu_1\nu_1\cdots \mu_s\nu_s}$. It transforms under $(s,0)\oplus (0,s)$ representation of $SO(4)$. The conformal dimension of both the irreps is $\Delta=2$. Since $C$ has dimension $2$, it will contribute $q^{2}\left[\chi_s(x_L)+\chi_s(x_R)\right]$ in the partition function. We also have to remove the contribution satisfied by the gauge conditions. In this case, the gauge conditions are given by
\begin{equation}
\begin{split}
&\mathcal{B}^{\mu_1\cdots\mu_s}\equiv\varepsilon^{\mu_1\nu_1\rho_1\sigma_1}\cdots \varepsilon^{\mu_s\nu_s\rho_s\sigma_s}\partial_{\nu_1}\cdots\partial_{\nu_s}C_{\rho_1\sigma_1\cdots \rho_s\sigma_s}=0~,\\&\partial_{\mu_1}\mathcal{B}^{\mu_1\cdots\mu_s}=0~.    
\end{split} 
\end{equation}
Now $\mathcal{B}$ has dimension $\Delta=s+2$ and it belongs to the irrep $\left(\frac{s}{2},\frac{s}{2}\right)$. So we need to subtract $q^{s+2}\chi_{\frac{s}{2}}(x_L)\chi_{\frac{s}{2}}(x_R)$. We also have to add $q^{s+3}\chi_{\frac{s-1}{2}}(x_L)\chi_{\frac{s-1}{2}}(x_R)$ to account for the second identity is satisfied. Here the tensor $\partial_{\mu_1}\mathcal{B}^{\mu_1\mu_2\cdots\mu_s}$ belongs to the irrep $\left(\frac{s-1}{2},\frac{s-1}{2}\right)$ and it has dimension $\Delta=s+3$. So the contribution for the single particle off-shell partition function is
\begin{equation}
z^{\rm off-shell}_s(k)=\frac{q^2\left[\chi_s(x_L)+\chi_s(x_R)-q^s\chi_{\frac{s}{2}}(x_L)\chi_{\frac{s}{2}}(x_R)+q^{s+1}\chi_{\frac{s-1}{2}}(x_L)\chi_{\frac{s-1}{2}}(x_R)\right]}{(1-qx_1)(1-qx^{-1}_1)(1-qx_2)(1-qx^{-1}_2)}~.     
\end{equation}
Also the equations of motions for the spin $s$ field is given by the generalized Bach equations. They are given by
\begin{equation}
\begin{split}
&B_{\mu_1\cdots\mu_s}\equiv\partial^{\nu_1}\cdots\partial^{\nu_s}C_{\mu_1\nu_1\cdots \mu_s\nu_s}=0~,\\&\partial^{\mu_1}B_{\mu_1\cdots\mu_s}=0~.     
\end{split}   
\end{equation}
So the on-shell contribution in the partition function can be obtained by the similar reasons. It is given by
\begin{equation}\label{zsons}
z_s^{\rm on-shell}(k)=\frac{q^2\left[q^s\chi_{\frac{s}{2}}(x_L)\chi_{\frac{s}{2}}(x_R)-q^{s+1}\chi_{\frac{s-1}{2}}(x_L)\chi_{\frac{s-1}{2}}(x_R)\right]}{(1-qx_1)(1-qx^{-1}_1)(1-qx_2)(1-qx^{-1}_2)}~.   
\end{equation}
So the complete single-particle partition function $z_s(k)$ can be written as
\begin{equation}
z_s(k)=z^{\rm off-shell}_s(k)-z_s^{\rm on-shell}(k)    
\end{equation}
which is the same as given in \eqref{partitions}. The complete partition function can now be computed by multi-particling the single-particle partition function.


In the limit of our interest \eqref{limit},  the partition function in the leading order is given by
\begin{equation}\label{logzchsbdy}
\ln Z^{\rm CHS}_s=\frac{\beta^2h_s(\beta)}{\nu_1\nu_2}+O\left(\frac{1}{\nu_i}\right)    
\end{equation}
where the residue function $h_s(\beta)$ is given by
\begin{equation}\label{reschs}
h_s(\beta)=\frac{1}{\beta^2}\sum_{k=1}^{\infty}\frac{\sinh\left(ks\beta\right)\sinh\left(k(s+1)\beta\right)}{2k^3\sinh^3\left(k\beta\right)}~.    
\end{equation}

We note the residue function  grows as $\frac{e^{2(s-1)k\beta}}{k^3}$ for the large $k$-terms in the sum. Therefore, the sum is only convergent for $s\leq 1$. This phenomenon can again be attributed to the presence of a negative twist state, with $\tau=-2(s-1)$, in the single particle partition function. Therefore, the theory of the conformal higher spins for $s\geq 2$, which is expected to violate the ANEC-type bound, also exhibits pathological behavior in the semi-universal limit. However, the traditional high-temperature expansion is completely insensitive to this behavior. We discuss more about this in section \ref{ansc}.
\subsection{Bulk determinant}\label{bulkchs}
It has been shown in \cite{Gupta:2012he,Beccaria:2014xda,Beccaria:2014jxa} that the partition function of conformal higher-spin fields (CHS) on $S^1_{\beta}\times S^3$ can be obtained from the bulk massless higher-spin partition function with both Dirichlet and Neumann boundary conditions. In this section, we briefly review the arguments of \cite{Beccaria:2014jxa,Beccaria:2014zma} and obtain the partition function of conformal higher-spin fields in the semi-universal limit \eqref{limit}.

The metric of the AdS space with unit radius is well known and is given by
\begin{align}
ds^2&=(1+r^2)d\tau^2+\frac{dr^2}{1+r^2}+r^2 d\Omega_3^2 ,
\end{align}
where $d\Omega_3^2$ is the metric on the unit three sphere.
For thermal AdS, we make the following identification
\begin{align}\label{thid}
(\tau,\phi_1,\phi_2)\sim
(\tau+\beta,\phi_1-i\beta\,\omega_1,\phi_2-i\beta\,\omega_2).
\end{align}
Let us consider the spin-$s$ field in the bulk $\phi_{\mu_1\mu_2\cdots \mu_s}$ in thermal AdS which admits the following mode decomposition:
\begin{align}
\Phi_{\mu_1\mu_2\cdots\mu_s}&=\sum_{n\in\mathbb Z}\sum_{L=0}^{\infty}\sum_{m_L=-j_L}^{j_L}
\sum_{m_R=-j_R}^{j_R}\Phi_{n,L;m_L,m_R}(r)\,
e^{i\omega_n\tau}\,
Y^{L}_{m_L,m_R}(\theta,\phi_1,\phi_2)_{\mu_1\cdots\mu_s},
\end{align}
where $Y^{L}_{m_L,m_R}(\theta,\phi_1,\phi_2)_{\mu_1\cdots\mu_s}$ is the spin-$s$ spherical harmonics on $S^3$.

The twisted boundary condition in \eqref{thid} implies that the Matsubara modes in the bulk are given by
\begin{align}
    \omega_n&=\frac{2\pi n}{\beta}-i\left(\omega_1 m_1+\omega_2 m_2\right),
\qquad m_1=m_L+m_R,\quad m_2=m_L-m_R .
\end{align}

The conformal higher spin fields (CHS) on the boundary are obtained by considering the ratio of alternate and standard quantization of bulk massless higher spin fields up to a finite possible zero mode contribution \cite{Beccaria:2014jxa,Beccaria:2014xda}.
\begin{align}\label{CHSpart}
Z^{\rm CHS}_{s}(S^1_\beta\times S^3)&=\frac{Z^{(-)}_{s,{\rm MHS}}(AdS_5)}{
Z^{(+)}_{s,{\rm MHS}}(AdS_5)}.
\end{align}
Here $Z^{(\pm )}_{s,{\rm MHS}}$ denotes the partition function obtained from two possible boundary conditions in AdS, namely Dirichlet $(+)$ and Neumann $(-)$. For a massless spin-$s$ field in the boundary $\Delta_+=s+2$ and $\Delta_-=2-s$.
The expression \eqref{CHSpart}, can be understood from the quantization of AdS/CFT. Here we briefly present the arguments of  \cite{Beccaria:2014jxa}. 
The near-boundary expansion of spin-$s$, bulk field $\Phi_s$ is given by \footnote{We denote the spin-$s$ bulk field $\Phi_{\mu_1\mu_2\cdots\mu_s}\equiv \Phi_s$.}
\begin{align}
    \Phi_s(z,x)\sim z^{\Delta_-}\, \phi_s(x)+z^{\Delta_+}\, \langle J_s(x)\rangle.
\end{align}
In the standard Dirichlet boundary condition (or $(+)$), we fix the boundary mode $\phi_s(x)$ and write the partition function as:
\begin{align}\label{mhs1loop}
Z^{(+)}_{s,\mathrm{MHS}}[\phi_s]
&=
\int_{\Phi_s|_{\partial \mathrm{AdS}}=\phi_s }
\mathcal{D}\Phi_s\,
e^{-S_{\mathrm{bulk}}[\Phi_s]} \nonumber\\
&=Z^{(+)}_{s,\mathrm{MHS}}[0]\,
\exp\left[
-\frac{1}{2}
\int_{S^1_\beta\times S^3}
d^4x\,\sqrt{g}\,
\phi_s\,\mathcal{O}^{\mathrm{CHS}}_s\,\phi_s
+\cdots
\right].
\end{align}

In the second line of \eqref{mhs1loop}, we have expanded the path integral at the one-loop and $\mathcal{O}^{\mathrm{CHS}}_s$ is the kinetic operator of the conformal higher spin field at the quadratic level. In the Neumann boundary condition (or $(-)$), we do not fix the boundary field $\phi_s(x)$, we rather integrate it over.
\begin{align}
    Z^{(-)}_{s,\mathrm{MHS}}&=\int [D\phi_s]\,Z^{(+)}_{s,\mathrm{MHS}}[\phi_s]\nonumber\\
    &=Z^{(+)}_{s,\mathrm{MHS}}[0]\,\int D[\phi_s]\exp\left[
-\frac{1}{2}
\int_{S^1_\beta\times S^3}
d^4x\,\sqrt{g}\,
\phi_s\,\mathcal{O}^{\mathrm{CHS}}_s\,\phi_s
\right]\nonumber\\
&=Z^{(+)}_{s,\mathrm{MHS}}\,\times  Z^{\rm CHS}_{s}(S^1_\beta\times S^3).
\end{align}
This provides the justification of \eqref{CHSpart} at the one-loop level.
The one-loop determinant for a massless  symmetric spin-$s$ field in
$AdS_5$ with a standard boundary condition is known, and it is given by \cite{David:2009xg,Gupta:2012he, Beccaria:2014jxa}
\begin{align}\label{dets}
Z^{(+)}_{s,{\rm MHS}}&=\left[\frac{\det\nolimits^{\rm TT}_{s-1}\left(-\nabla^2+(s+2)(s-1)\right)}{\det\nolimits^{\rm TT}_{s}\left(-\nabla^2+((s+2)(s-2)-s\right)}\right]^{1/2}.
\end{align}
Here, the denominator comes from the transverse-traceless spin-$s$ determinant,
whereas the numerator is the corresponding transverse-traceless spin-$(s-1)$ determinant coming from the ghost contribution.

We now compute \eqref{CHSpart} from the bulk. We label a spatial normal mode by its energy $E_A$ and angular quantum
numbers $m_{1A}, m_{2,A}$. Here $A=\pm$ denotes the two possible
quantizations, namely the Dirichlet and the Neumann. We also define
\begin{align}\label{thetaA}
    \theta_A& = -i\beta\left(\omega_1\, m_{1,A}+\omega_2\, m_{2,A}\right).
\end{align}
The contribution of this mode to the bosonic determinant is
\begin{align}
    \ln Z_A
    =
    -\frac12
    \sum_{n\in\mathbb Z}
    \ln\left[
        \left(\frac{2\pi n-\theta_A}{\beta}\right)^2
        +E_A^2
    \right].
\end{align}
Using the standard Matsubara sum, we find
\begin{align}
\sum_{n\in\mathbb Z}
\ln\left[
        \left(\frac{2\pi n-\theta}{\beta}\right)^2
        +E^2
    \right]
=
\beta E
+
\ln\left(1-e^{-\beta E+i\theta}\right)
+
\ln\left(1-e^{-\beta E-i\theta}\right)
+\sum_{n\in \mathbb{Z}}\ln \frac{1}{\beta^2},
\end{align}
Dropping the Casimir term, and also the $E_A$, $\theta$-independent divergent term, we get
\begin{align}
    \ln Z_{A,\rm th}&  =
    -\ln\left(
        1-e^{-\beta E_A+i\theta_A}
    \right)\nonumber\\
    &=  \sum_{k=1}^{\infty}
    \frac1k\,
    e^{-k\beta E_A}
    e^{ik\theta_A}
\end{align}
In the character sum below, we will include the modes with opposite angular momenta separately.
Summing over all spatial modes, and using the definition of $\theta_A$ in \eqref{thetaA} we obtain
\begin{align}
    \ln Z
    =
    \sum_{k=1}^{\infty}
    \frac1k
    \sum_{A=\pm}
    e^{-k\beta E_A}
    e^{k\beta(\omega_1m_{1,A}+\omega_2m_{2,A})}.
\end{align}

For a massless spin-\(s\) field in thermal \(AdS\), the physical
transverse-traceless spin-\(s\) modes are accompanied by the
transverse-traceless spin-\((s-1)\) ghost modes. Therefore, the standard
quantization contribution takes the form
\begin{align}
\ln Z^{(+)}_{s,{\rm MHS}}
&=
\sum_{k=1}^{\infty}
\frac1k
\left[
\sum_{a\in{\rm TT}_s}
e^{-k\beta E_{s,a}}
e^{k\beta(\omega_1m_{1,a}+\omega_2m_{2,a})}
\right.
\left.
-
\sum_{b\in{\rm TT}_{s-1}}
e^{-k\beta E_{{\rm gh},b}}
e^{k\beta(\omega_1m_{1,b}+\omega_2m_{2,b})}
\right].
\end{align}
Here $\rm{TT}_s$ denotes the representation of a transverse traceless spin-$s$ field. We can perform the sum over transverse-traceless modes and rewrite the partition function in terms of 
$SU(2)_L\times SU(2)_R$ characters. Using the standard 
quantization of a massless spin-$s$ field in AdS$_5$, the physical
spin-$s$ modes have lowest energy
\begin{align}
    \Delta_{+,s}=s+2,
\end{align}
whereas the ghost spin-$(s-1)$ modes have lowest energy\footnote{We can understand this easily by looking at the mass term of the determinant in \eqref{dets}.}
\begin{align}
    \Delta_{+,{\rm gh}}=s+3 .
\end{align}
Therefore the corresponding one-particle partition function takes the form
\begin{align}\label{zplus}
z^{(+)}_{s,{\rm MHS}}
&=\frac{
q^{s+2}\,\chi_{\frac{s}{2}}(x_L)\chi_{\frac{s}{2}}(x_R)
-
q^{s+3}\,\chi_{\frac{s-1}{2}}(x_L)\chi_{\frac{s-1}{2}}(x_R)
}
{
(1-qx_1)(1-qx_1^{-1})(1-qx_2)(1-qx_2^{-1})
},
\end{align}

$x_L, x_R$ are defined by the following relations
\begin{align}
    x_L=\left(x_1x_2\right)^{\frac{1}{2}},
    \qquad
    x_R=\left(\frac{x_1}{x_2}\right)^{\frac{1}{2}}.
\end{align}
Note that, \eqref{zplus} is the same expression as the on-shell single-particle partition function as given in \eqref{zsons}.
Thus the partition function of the modes with the standard quantization is obtained by multiparticling the single-particle partition function
\begin{align}\label{fullzplus}
\ln Z^{(+)}_{s,{\rm MHS}}=
\sum_{k=1}^{\infty}
\frac{1}{k}\,
z^{(+)}_{s,{\rm MHS}}
(q^k,x_1^k,x_2^k).
\end{align}
For the alternative quantization or with the Neumann boundary condition, the boundary value $\phi_s$ is not a lowest-weight bulk particle with the dimension , $\Delta =2-s$, and this is obvious because the particle should not have negative energy for $s>2$. Therefore, it is a conformal gauge field at the boundary, and under the gauge transformation, it transforms as
\begin{align}
    \delta\phi_{\mu_1\cdots \mu_s}=\partial_{\mu_1} \xi_{\mu_2\cdots \mu_s}-\rm {trace}.
\end{align}
Here $\xi_{\mu_2\cdots\mu_s}$ is the gauge parameter with the dimension $1-s$. Therefore, the contribution from this module is given by
\begin{align}
    \hat{z}^{-}_{s,\rm MHS}&=\frac{
q^{2-s}
\chi_{\frac{s}{2}}(x_L)\chi_{\frac{s}{2}}(x_R)
-
q^{1-s}
\chi_{\frac{s-1}{2}}(x_L)\chi_{\frac{s-1}{2}}(x_R)
}
{
(1-qx_1)(1-qx_1^{-1})(1-qx_2)(1-qx_2^{-1})
}.
\end{align}
But, we also have to also consider the contribution from the conformal killing tensors which are basically the residual gauge degrees of freedom.
The finite character corresponding to this conformal killing tensor is given by 
\begin{align}
&\sigma_s(q,x_1,x_2)\\
&=
\frac{
q^2\left[\chi_s(x_L)+\chi_s(x_R)\right]
-
\left(q^{s+2}+q^{2-s}\right)
\chi_{\frac{s}{2}}(x_L)\chi_{\frac{s}{2}}(x_R)
+
\left(q^{s+3}+q^{1-s}\right)
\chi_{\frac{s-1}{2}}(x_L)\chi_{\frac{s-1}{2}}(x_R)
}
{
(1-qx_1)(1-qx_1^{-1})(1-qx_2)(1-qx_2^{-1})
}~.
\end{align}
In the non-rotating limit, $x_1=x_2=1$, it reproduces the finite character corresponding to the conformal killing tensor, given in \cite{Beccaria:2014jxa}.

The single-particle partition function corresponding to the alternative quantization is given by
\begin{align}\label{zminus}
    z^{(-)}_{s,\rm MHS}&=\hat{z}^{-}_{s,\rm MHS}+\sigma_s(q)\nonumber\\
    &=\frac{
q^2\left[
\chi_s(x_L)+\chi_s(x_R)
-q^s\chi_{\frac{s}{2}}(x_L)\chi_{\frac{s}{2}}(x_R)
+q^{s+1}\chi_{\frac{s-1}{2}}(x_L)\chi_{\frac{s-1}{2}}(x_R)
\right]
}
{
(1-qx_1)(1-qx_1^{-1})(1-qx_2)(1-qx_2^{-1})
}~.
\end{align}
Therefore, the partition function corresponding to the alternative quantization is obtained by multi-particling the single-particle partition function.
\begin{align}\label{fullzm}
    \ln Z^{(-)}_{s,{\rm MHS}}=
\sum_{k=1}^{\infty}
\frac{1}{k}\,
z^{(-)}_{s,{\rm MHS}}
(q^k,x_1^k,x_2^k)~.
\end{align}
The boundary conformal higher-spin partition function is therefore
\begin{align}\label{logzchs}
\ln Z^{\rm CHS}_{s}
&=\ln Z^{(-)}_{s,{\rm MHS}}-\ln Z^{(+)}_{s,{\rm MHS}},
\end{align}
where $Z^{(+)}_{{\rm sp},s,{\rm MHS}}$ and $Z^{(-)}_{{\rm sp},s,{\rm MHS}}$ are given in \eqref{fullzplus} and \eqref{fullzm}.

We find that the partition function computed from the bulk one-loop determinant with the Dirichlet and Neumann boundary conditions agrees with the partition function of the conformal higher spins at the boundary \cite{Beccaria:2014jxa, Tseytlin:2013jya}.

We now consider the semi-universal limit \eqref{limit} of the partition function, which  indeed agrees with \eqref{logzchsbdy} and the residue function $h_s(\beta)$ given in \eqref{reschs}. In the following section, we will show that the residue function $h_s(\beta)$ captures the violation of the ANEC bound \cite{Hofman:2008ar, Cordova:2017dhq} and provides the semi-universal limit as a diagnostic of the ANEC violation. But the traditional high temperature expansion of the partition function of the conformal higher spins is completely insensitive to the violation of ANEC.
\section{Semi-universality as a diagnostic of ANEC-type violation}\label{ansc}
In this section, we compare two different limits of the partition function. First,
we consider the small-$\beta$ limit, $\beta\to 0$. We then consider the
semi-universal limit $\nu_i\to 0$ at fixed $\beta$.
The important point is that these two limits do not necessarily commute. 

If the small-$\beta$ expansion is taken first, the leading terms are governed by ordinary zeta-function sums, such as $\sum_{k=1}^{\infty}\frac{1}{k^4}$, and the partition functions exhibit the expected high-temperature behavior, which is $\beta^{-3}$. In contrast, if the semi-universal limit $\nu_i \to 0$, is taken first at fixed $\beta$, higher-derivative theories with negative-twist states, together with CHS fields with $s\geq 2$, develop an ill-defined partition function.  More precisely, for CHS, the summand itself grows exponentially at large $k$,
$
\sim e^{2(s-1) k\beta}$
, so that the defining $k$-sum is divergent for $s>1$. This exponential growth reflects the presence of negative-twist states in the corresponding  ANEC violating spectrum. 

Conversely, for theories whose spectrum contains only positive-twist states, the residue function $h(\beta)$ remains well defined in the semi-universal limit. 
Theories that violate the naive unitarity bounds but do not contain negative-twist states
can still have well-defined partition functions and well-behaved residue functions in the
semi-universal limit. In contrast, theories that obey the naive unitarity bounds but contain
negative-twist states associated with ANEC-type bound violations can exhibit ill-defined
partition functions and pathological behavior in the semi-universal limit.

Thus, the semi-universal limit provides a sharp diagnostic for the presence of negative-twist states: when such states are present, the large-$k$ terms in the residue function can grow exponentially and render $h(\beta)$ ill defined; when they are absent, the residue remains well behaved.

In the table below, we show the properties of the residue function in two different limits: 
\begin{enumerate}
    \item First, in the small $\beta$-expansion.
    \item  Second, in the small $\nu$-expansion, with fixed $\beta$.
\end{enumerate}  
We present the residue functions of the leading singularities of the partition functions for the conformal fields studied in this paper, emphasizing their behavior in these two non-commuting limits.
\begin{table}[H]
\centering
\small
\renewcommand{\arraystretch}{1.8}
\begin{tabular}{|c|c|c|}
\hline
Field 
&
Small-\(\beta\) first
&
Small-\(\nu_i\) first, fixed \(\beta\)
\\
\hline

\(\phi\)
&
\(\displaystyle
\frac{1}{\nu_1\nu_2}
\frac{1}{2\beta}
\sum_{k=1}^{\infty}\frac{1}{k^4}
\)
&
\(\displaystyle
\frac{1}{\nu_1\nu_2}
\sum_{k=1}^{\infty}
\frac{e^{-k\beta}}{k^3(1-e^{-2k\beta})}
\)
\\
\hline

\(\phi_{4\partial}\)
&
\(\displaystyle
\frac{1}{\nu_1\nu_2}
\frac{1}{\beta}
\sum_{k=1}^{\infty}\frac{1}{k^4}
\)
&
\(\displaystyle
\frac{1}{\nu_1\nu_2}
\sum_{k=1}^{\infty}
\frac{\operatorname{coth}(k\beta)}{k^3}
\)
\\
\hline

\(\psi\)
&
\(\displaystyle
\frac{1}{\nu_1\nu_2}
\frac{2}{\beta}
\sum_{k=1}^{\infty}
\frac{(-1)^{k+1}}{k^4}
\)
&
\(\displaystyle
\frac{1}{\nu_1\nu_2}
\sum_{k=1}^{\infty}
(-1)^{k+1}
\frac{2e^{-k\beta/2}}{k^3(1-e^{-k\beta})}
\)
\\
\hline

\(\psi_{3\partial}\)
&
\(\displaystyle
\frac{1}{\nu_1\nu_2}
\frac{6}{\beta}
\sum_{k=1}^{\infty}
\frac{(-1)^{k+1}}{k^4}
\)
&
\(\displaystyle
\frac{1}{\nu_1\nu_2}
\sum_{k=1}^{\infty}
\frac{2(-1)^{k+1}}{k^3}
\frac{
e^{k\beta/2}+e^{-k\beta/2}+e^{-3k\beta/2}
}
{1-e^{-k\beta}}
\)
\\
\hline

\(A_\mu\)
&
\(\displaystyle
\frac{1}{\nu_1\nu_2}
\frac{1}{\beta}
\sum_{k=1}^{\infty}\frac{1}{k^4}
\)
&
\(\displaystyle
\frac{1}{\nu_1\nu_2}
\sum_{k=1}^{\infty}
\frac{\operatorname{coth}(k\beta)}{k^3}
\)
\\
\hline
\(A^{6d}_{M,\,4\partial}\)
&
\(\displaystyle
\frac{1}{\nu_{1}\nu_{2}\nu_{3}}\,
\frac{2}{\beta^{2}}
\sum_{k=1}^{\infty}\frac{1}{k^{6}}
\)
&
\(\displaystyle
\frac{1}{\nu_{1}\nu_{2}\nu_{3}}
\sum_{k=1}^{\infty}
\frac{2\coth^{2}(k\beta)}{k^{4}}
\)
\\
\hline
\(\psi_\mu^{\rm Weyl}\)
&
\(\displaystyle
\frac{1}{\nu_1\nu_2}
\frac{8}{\beta}
\sum_{k=1}^{\infty}
\frac{(-1)^{k+1}}{k^4}
\)
&
\(\displaystyle
\frac{1}{\nu_1\nu_2}
\sum_{k=1}^{\infty}
\frac{4(-1)^{k+1}\cosh^2(k\beta)}
{k^3\sinh\left(\frac{k\beta}{2}\right)}
\)
\\
\hline

\(h_{\mu\nu}^{\rm Weyl}\)
&
\(\displaystyle
\frac{1}{\nu_1\nu_2}
\frac{3}{\beta}
\sum_{k=1}^{\infty}\frac{1}{k^4}
\)
&
\(\displaystyle
\frac{1}{\nu_1\nu_2}
\sum_{k=1}^{\infty}
\frac{
\operatorname{coth}(k\beta)
\left(3+4\sinh^2(k\beta)\right)
}{k^3}
\)
\\
\hline

\(\Phi_s\)
&
\(\displaystyle
\frac{1}{\nu_1\nu_2}
\frac{s(s+1)}{2\beta}
\sum_{k=1}^{\infty}\frac{1}{k^4}
\)
&
\(\displaystyle
\frac{1}{\nu_1\nu_2}
\sum_{k=1}^{\infty}
\frac{
\sinh(sk\beta)\sinh((s+1)k\beta)
}
{2k^3\sinh^3(k\beta)}
\)
\\
\hline

\end{tabular}
\caption{
The partition function in the small-\(\beta\) expansion vs the semi-universal small-\(\nu_i\) limit at fixed \(\beta\).
}
\label{tab2}
\end{table}

Note that the twist
$
\tau=E-J_1-J_2
$
is not guaranteed to be positive from the standard representation-theoretic unitarity bounds alone. Primary operators in a four-dimensional CFT are labeled by the highest weight
$
(E,j_L,j_R)=\left(E,\frac{j_1+j_2}{2},\frac{j_1-j_2}{2}\right),
$
and for $j_1\geq j_2\geq0$, the unitarity bounds take the form \cite{Mack:1975je,Minwalla:1997ka}
\begin{equation}
E \geq
\begin{cases}
\max(j_1,j_2)+2, & \text{if } j_1\neq j_2,\\[4pt]
j+1, & \text{if } j_1=j_2=j>0,\\[4pt]
0, & \text{if } j_1=j_2=0 .
\end{cases}
\end{equation}
From the unitarity bound, the minimal twist can be negative as well since $J_i$ runs from $-j_i$ to $j_i$.
However, CFTs are known to obey the stronger constraint
\begin{equation}
E \geq j_1+j_2,
\qquad \text{if } \min(j_1,j_2)>2 .
\end{equation}
as a consequence of ANEC-based arguments \cite{Cordova:2017dhq,Hofman:2008ar}. Thus, the positivity of the twist is tied to a stronger dynamical constraint in conformal field theory, rather than to the basic representation-theoretic unitarity bounds alone.

The connection between semi-universality and the ANEC bound can be understood simply, from the definition of the thermal trace in  \eqref{partdef}. The semi-universal limit, probes into the large-angular momentum regime, and the sign of the twist becomes important here. The twist $\tau=E-J_1-J_2$ is only positive when the ANEC bound is satisfied; therefore, a negative twist state contributes as $e^{+\beta|\tau|}$ to the single-particle partition function, which also violates the ANEC bound \cite{Hofman:2008ar,Cordova:2017dhq}. Therefore, when we form the multi-particle partition function for a negative twist state, the partition function diverges.

In contrast, the ordinary high-temperature expansion does not detect this pathology. In the high temperature limit, the partition function is dominated by many derivatives of all sorts acting on every field.
 For this reason, the typical `letter' that contributes to the high temperature limit  is of positive twist, even though the bare (derivative free) single particle partition function is of negative twist. In the semi-universal limit, however, 
the typical `letter' still has many derivatives, but all with zero twist. The number of positive twist derivatives is small, 
and so this cannot mask the pathology of the negative twist field.
Therefore, the semi-universal limit provides a sharp diagnostic of whether the spectrum of a conformal field theory is compatible with the ANEC bound \cite{Hofman:2008ar,Cordova:2017dhq}.

\section{Discussion}
In this paper, we study the partition functions of exotic conformal field theories, focusing on conformal higher-derivative fields and conformal higher-spin theories in the semi-universal limit \eqref{limit}. We show that these partition functions indeed develop the universal pole structure conjectured in \cite{Anand:2025mfh}, while the corresponding residue functions remain theory-dependent. We then analyze the properties of the residue function associated with the leading singularity in this limit. In particular, we find that conformal higher-spin fields with $s\geq 2$, which are expected to violate the ANEC bound, also exhibit pathological behavior in the semi-universal limit. However, in the traditional high- temperature expansion, the residue functions are well behaved, and therefore, semi-universality provides an important diagnostic of the ANEC bounds of the conformal field theories.

Beyond the exotic free conformal field theories studied in this paper, it would be interesting to investigate thermal partition functions of interacting conformal field theories, such as $O(N)$ vector models \cite{David:2024pir, David:2025tqn} and matter Chern-Simons theories in three dimensions. It was also conjectured in \cite{Anand:2025mfh} that the residue function of the leading singularity of the partition function contains non-perturbative terms in the small-$\beta$ expansion, for example, terms of the form $e^{-a/\beta}$, only in free theories. It would therefore be useful to study the corresponding residue functions in interacting theories and test this prediction of \cite{Anand:2025mfh}.

It would also be interesting to explore the entanglement entropy of these higher-spin fields from the thermal partition function on the branched sphere \cite{Mukherjee:2021rri}, in the presence of global charges, and to understand how they behave when the chemical potentials are tuned to their extreme values. It is well known that the entanglement entropy of conformal fields across a half-space can be obtained using the thermal partition function on the hyperbolic cylinder \cite{Casini:2011kv}. However, the hyperbolic-cylinder partition functions do not capture the contribution coming from the `edge modes' for higher-spin fields, whereas the sphere partition functions compute both the bulk contribution and the contribution from the edge modes \cite{Anninos:2020hfj,Anninos:2026hia,David:2022jfd}.

It would also be interesting to evaluate the bulk one-loop determinant using quasinormal modes \cite{Denef:2009kn}. Presumably, the quasinormal-mode character captures only the bulk one-loop determinant \cite{Anninos:2020hfj,Anninos:2026hia}. The corresponding edge character can then be obtained by dividing the complete one-loop determinant, evaluated using Euclidean methods, by the bulk-character partition function. It would be interesting to study the properties of this edge-character partition function for these exotic theories in the bulk and to understand whether the edge partition function admits a representation as a local path integral on a codimension-two surface \cite{Anninos:2020hfj, Law:2025yec, Law:2026tuk}; see also \cite{Mukherjee:2025xlt, Mukherjee:2024nhx}.

We would also like to explore semi-universality for conformal supergravity in AdS$_5$ \cite{Beccaria:2014xda} and in higher dimensions. Another interesting direction would be to test vectorial AdS$_5$/CFT$_4$ duality for the spin-1 boundary theory \cite{Beccaria:2014zma} in the semi-universal limit.
\subsection*{Acknowledgement} We thank Shiraz Minwalla for discussions and  valuable suggestions on the manuscript. We also thank Justin David for the comments on the manuscript. The research of J.M. and P.R. is supported by the Department of Atomic Energy, Government of India and the Infosys Endowment for the study of the Quantum Structure of Spacetime. 
\appendix

\section{Conformal higher derivative vector field on $S^1_{\beta}\times S^5$: mode sum approach}\label{app1}

In this appendix, we provide the details of the spectral mode-sum approach to compute the partition function of the Weyl invariant higher derivative vector field on $S^1_{\beta}\times S^5$.
We begin with the gauge fixed partition function \cite{Beccaria:2015uta, Beccaria:2017dmw}:
\begin{equation}
\ln Z_1=-\frac{1}{2}\ln\det\mathcal{O}_1-\frac{1}{2}\ln\det\mathcal{O}_2    
\end{equation}
where $\mathcal{O}_1=-(\partial_0+1)^2-\nabla_{S^5}^2+5$ and $\mathcal{O}_2=-(\partial_0-1)^2-\nabla_{S^5}^2+5$.

Using the Schwinger parameter, we can express the first term as
\begin{equation}
-\frac{1}{2}\ln\det\mathcal{O}_1=\frac12 \int_0^\infty \frac{ds}{s}\,
{\rm Tr}\left(e^{-s\mathcal O_1}\right).    
\end{equation}
Twisted boundary conditions \eqref{twistedbcvector1} implies that the Matsubara frequencies are
\begin{equation}
 \omega_n=\frac{2\pi n}{\beta}-i\mu   
\end{equation}
where $\mu(m_1,m_2,m_3)$ are given by
\begin{equation}
\mu(m_1,m_2,m_3)=\omega_1 m_1+\omega_2m_2+\omega_3m_3~.    
\end{equation}
The eigenvalue of the operator $\mathcal{O}_1$ is given by\footnote{For the transverse Vector spherical harmonics, we have $-\nabla^2_{S^5}+5\to (\ell+3)^2$.}
\begin{equation}
\lambda^{(1)}_{n,\ell}=\left(\frac{2\pi n}{\beta}-i\left(\mu+1\right)\right)^2+\left(\ell+3\right)^2    
\end{equation}
where $m_i$ are the eigenvalues of $J_i$ for $i=1,2,3$.\\\\
Hence, we can express the first term of $\ln Z_1$ (after manipulations with the Poisson resummation and neglecting the Casimir energy) as given by
\begin{equation}
\begin{split}
-\frac{1}{2}\ln\det\mathcal{O}_1=\frac{1}{2}
\sum_{\ell=0}^{\infty}\sum_{m_1,m_2,m_3}
\sum_{k\in\mathbb Z\setminus\{0\}}
{\rm mult}(m_1,m_2,m_3)\frac{1}{|k|}
e^{k\beta(\mu+1)}
e^{-\beta|k|(\ell+3)} .    
\end{split}    
\end{equation}
where ${\rm mult}(m_1,m_2,m_3)$ is the multiplicity of the weight $(m_1,m_2,m_3)$.\\Similarly, for the operator $\mathcal{O}_2$, we have
\begin{equation}
-\frac{1}{2}\ln\det\mathcal{O}_2=\frac{1}{2}
\sum_{\ell=0}^{\infty}\sum_{m_1,m_2,m_3}
\sum_{k\in\mathbb Z\setminus\{0\}}
{\rm mult}(m_1,m_2,m_3)\frac{1}{|k|}
e^{k\beta(\mu-1)}
e^{-\beta|k|(\ell+3)} .    
\end{equation}
So the full log of the partition function is
\begin{equation}
\begin{split}
\ln Z_1&=
\sum_{\ell=0}^{\infty}\sum_{m_1,m_2,m_3}
\sum_{k\in\mathbb Z\setminus\{0\}}
{\rm mult}(m_1,m_2,m_3)\frac{1}{|k|}
e^{k\beta\mu}\cosh(k\beta)
e^{-\beta|k|(\ell+3)}\\&=\sum_{k\in\mathbb{Z}\setminus\{0\}}\sum_{\ell=0}^{\infty}\frac{1}{|k|}\cosh(k\beta)e^{-\beta|k|(\ell+3)}\\&~~~~~~~~~~~~~~~~~~~~~~~\times\sum_{m_1,m_2,m_3}{\rm mult}(m_1,m_2,m_3)e^{k\beta(\omega_1 m_1+\omega_2m_2+\omega_3m_3)}~.   
\end{split}    
\end{equation}
We can see that $m_1,m_2,m_3$ sum can be expressed as $SO(6)$ Character $\chi^{SO(6)}_{(\ell+1,1,0)}(x_1,x_2,x_3)$. We have used the following property of the Character
\begin{equation}
\chi^{SO(6)}_{(\ell+1,1,0)}(x_1,x_2,x_3)=\chi^{SO(6)}_{(\ell+1,1,0)}(x_1^{-1},x_2^{-1},x_3^{-1})~.    
\end{equation}
So the log of the partition function can be expressed as
\begin{equation}
\ln Z_1=\sum_{k=1}^{\infty}\frac{2\cosh(k\beta)e^{-3k\beta}}{k}\sum_{\ell=0}^{\infty}e^{-k\beta\ell}\chi^{SO(6)}_{(\ell+1,1,0)}\big(e^{k\beta\omega_1},e^{k\beta\omega_2},e^{k\beta\omega_3}\big)~.    
\end{equation}
The expression for the Character $\chi^{SO(6)}_{(\ell+1,1,0)}(x_1,x_2,x_3)$ is given by
\begin{equation}
\begin{split}
 &\chi^{SO(6)}_{(\ell+1,1,0)}(x_1,x_2,x_3)\\&=\frac{1}{(y_1-y_2)(y_2-y_3)(y_3-y_1)}\\&\times\Big[\left(x_1^{\ell+3}+x_1^{-(\ell+3)}\right)(y_2^2-y_3^2)+\left(x_2^{\ell+3}+x_2^{-(\ell+3)}\right)(y_3^2-y_1^2)+\left(x_3^{\ell+3}+x_3^{-(\ell+3)}\right)(y_1^2-y_2^2)\Big]   
\end{split}    
\end{equation}
where $y_i=x_i+x_i^{-1}$ for $i=1,2,3$. We will use the following generating function of the Character to perform the $\ell$ sum:
\begin{equation}
\begin{split}
&\sum_{\ell=0}^{\infty}t^\ell \chi^{SO(6)}_{(\ell+1,1,0)}(x_1,x_2,x_3)\\&=\frac{1}{(1-tx_1)(1-tx^{-1}_1)(1-tx_2)(1-tx^{-1}_2)(1-tx_3)(1-tx^{-1}_3)}\\&~~~~~~\Big[3+y_1y_2+y_2y_3+y_3y_1-t(y_1y_2y_3+3y_1+3y_2+3y_3)+t^2(4+y_1y_2+y_2y_3+y_3y_1)\\&~~~~~~~~~~~~~~~-t^3(y_1+y_2+y_3)+t^4\Big]~.    
\end{split}  
\end{equation}
So the final result for the log of the partition function is given by
\begin{equation}
\begin{split}
\ln Z_1=\sum_{k=1}^{\infty}\frac{1}{k}z_1(k)    
\end{split}    
\end{equation}
where the single particle partition function for the higher derivative vector is 
\begin{equation}
\begin{split}
&z_1(k)\\&=\frac{e^{-2k\beta}\left(1+e^{-2k\beta}\right)}{\prod_{i=1}^{3}\left(1-e^{-k\beta(1-\omega_i)}\right)\left(1-e^{-k\beta(1+\omega_i)}\right)
 }\\&\times\Big[3+y_1(k)y_2(k)+y_2(k)y_3(k)+y_3(k)y_1(k)-e^{-k\beta}\big(y_1(k)y_2(k)y_3(k)+3y_1(k)+3y_2(k)+3y_3(k)\big)\\&~~~~~+e^{-2k\beta}\big(4+y_1(k)y_2(k)+y_2(k)y_3(k)+y_3(k)y_1(k)\big)-e^{-3k\beta}\big(y_1(k)+y_2(k)+y_3(k)\big)+e^{-4k\beta}\Big]    
\end{split}    
\end{equation}
where $y_i(k)$ is defined as
\begin{equation}
y_i(k)\equiv 2\cosh(k\beta\omega_i)~,\qquad i=1,2,3~.    
\end{equation}

\bibliographystyle{JHEP}
\bibliography{biblio}

\end{document}